\documentclass[journal]{IEEEtran}

\usepackage{cite}
\usepackage{amsmath,amssymb,amsfonts}
\usepackage{graphicx}
\usepackage{booktabs}
\usepackage{siunitx}
\usepackage{url}
\usepackage[hidelinks]{hyperref}

\begin{document}

\title{An Absolute Acoustic Readout of Plate Wave Speed for Bullion Authentication,
Using a Measured Fill-Factor Correction for Struck Coins}

\author{Sylvia~Dong and Takyiu~Liu%
  \thanks{Preprint, October 2026.}%
  \thanks{S. Dong is with Lynbrook High School, San Jose, CA, USA
  (e-mail: sylviad0407@gmail.com).}%
  \thanks{T. Liu is with West Valley College, Saratoga, CA, USA
  (e-mail: takyiu.liu@westvalley.edu).}%
}

\markboth{Preprint}{Dong and Liu: Absolute Acoustic Readout of Plate Wave Speed}
\maketitle

\begin{abstract}
Acoustic ``ping'' testing of bullion is empirical: it compares a coin's ringing
frequency with frequencies recorded for the same coin type, so it cannot assess a
specimen absent from the reference library. This work replaces the
comparison with an absolute measurement, computed from the measured resonance and
geometry through a physical model rather than retrieved from a library.
Free-plate flexural resonance is calibrated on five machined aluminium discs
($a/t=2.58$--$10.75$), recovering an empirical finite-thickness coefficient
$\beta=0.895\pm0.047$ that reduces the worst-case departure from thin-plate theory from
\SI{8.9}{\percent} to \SI{0.75}{\percent}. A fill factor, the ratio of a coin's
displaced volume to its nominal cylinder, supplies the effective thickness that relief
hides from calipers, extending the calibration to struck coins. The readout is the plate
wave speed $c_{p}=\sqrt{E/\rho(1-\nu^{2})}$, a material constant, conditioned on an
assumed Poisson's ratio. On six struck 0.999 copper rounds not used in the
calibration it recovers a Young's modulus of $119.1\pm6.7$\,\si{\giga\pascal},
inside the accepted \SIrange{110}{128}{\giga\pascal}. The recovered wave speeds form
distinct observed groups across the measured specimens, with the strongest independent
validation provided by these copper rounds. The $(3,0)/(2,0)$ mode ratio falls below
its thin-plate value in thirteen of sixteen specimens (one-sided sign test, $p=0.011$),
as finite-thickness theory predicts. Weighing displaced mass to
\SI{0.1}{\milli\gram} rather than \SI{0.1}{\gram} cut per-specimen volume uncertainty
from \SIrange{1.6}{30}{\percent} to \SIrange{0.5}{2.6}{\percent} and shrank an apparent
\SI{11}{\percent} design-dependent bias to a \SI{4.7}{\percent} residual on one deeply
relieved silver design. The readout is now limited by tabulated elastic constants and
the assumed Poisson's ratio, not by the acoustics, which are reproducible to
\SI{0.15}{\percent}.
\end{abstract}


\section{Introduction}
\IEEEPARstart{S}{ilver} bullion is an attractive counterfeiting target because its
value derives almost entirely from metal content rather than from craftsmanship or
provenance. A convincing fake need only match a coin's diameter, thickness, mass and
appearance---a specification that is published, standardised, and straightforward to
approximate in base metal.

The established non-destructive tests each stop somewhere specific. Magnet tests are
free and defeat any ferrous-core fake, but say nothing about a non-ferrous one.
Specific gravity is genuinely informative, yet base-metal cores chosen to match the
density of gold or silver make it unsuitable as a stand-alone test~\cite{bullionguide};
a correctly sized tungsten core passes it. X-ray fluorescence reads elemental
composition accurately but only at the surface, to a depth of order tens of
micrometres in a dense metal, so a plating thicker than that returns a clean reading on
a base-metal interior. That limitation is structural
rather than a matter of instrument cost: XRF is a surface technique being asked a
bulk question. Ultrasonic pulse-velocity testing does address the bulk, and rests on
the observation this work also exploits---that substitute metals matching a precious
metal's \emph{density} do not match its \emph{elasticity}. Gold and tungsten differ by
\SI{0.3}{\percent} in density, yet the elastic constants of
Table~\ref{tab:handbook} put their plate wave speeds at \SI{2240}{\meter\per\second}
and \SI{4810}{\meter\per\second}: a factor of $2.1$ on the quantity this work
recovers. Pulse-velocity testing exploits the same contrast, but requires a contact
transducer, couplant, and a dedicated instrument.

Acoustic resonance offers the same bulk sensitivity from a microphone. Dao
\emph{et al.}~\cite{dao2022} built an electromagnetic acoustic transducer system and
separated four possible counterfeits from thirty authentic coins using a threshold on the
\SIrange{16.9}{17.5}{\kilo\hertz} band, with repeatability of $\pm$\SI{12}{\hertz}.
Several consumer applications now implement the same idea; the most developed is
optimised for a fixed list of supported coins---twenty-three silver coins when last
accessed---and its vendor notes that privately minted rounds, whose specifications
vary, are tested less reliably~\cite{pingcoin}.

Both share one structural limitation: neither measures a material property. Both
measure a \emph{frequency} and compare it to frequencies previously observed for that
exact object. A frequency is not a property of silver---it is a property of this
piece of silver, at this diameter and this thickness. Change the geometry and the
frequency changes, so the reference library must be rebuilt for every coin type.
Coverage is therefore quoted in units of supported coins.

The same structural point separates this work from the wider non-destructive
literature. Impulse excitation as standardised in ASTM E1876~\cite{astm1876} does
recover absolute elastic constants, but only for slender bars, rods and flat discs of
accurately known, uniform geometry; a struck coin is a disc whose thickness is neither
uniform nor measurable by calipers, which is why the standard has not been applied to
bullion. Eddy-current
and electrical-conductivity instruments interrogate the bulk and are sold to
dealers, but they respond to conductivity rather than to elasticity, and a
conductivity match is easier to engineer than an elastic one. Image-based
methods---three-dimensional height-map analysis of the relief, for
example---authenticate the \emph{die} rather than the metal, and are therefore
orthogonal to the question asked here. What is missing across all of these is a
bulk elastic measurement on an object of unknown effective geometry, obtained
without contact instrumentation.

This paper removes the library. Section~\ref{sec:theory} sets out the plate theory and
the two obstacles that have kept acoustic coin testing empirical.
Section~\ref{sec:method} describes the apparatus and the analysis pipeline.
Section~\ref{sec:results} calibrates the method on aluminium, validates it on copper
and silver, and applies it to five bullion-like rounds of unverified composition. Section~\ref{sec:discussion}
quantifies where the method breaks down.

\section{Theory}\label{sec:theory}

For a free circular plate, Kirchhoff thin-plate theory gives the natural frequencies
as
\begin{equation}
  f(n,s)=\frac{\lambda^{2}(n,s)}{2\pi a^{2}}
         \sqrt{\frac{E t^{2}}{12\rho\left(1-\nu^{2}\right)}} ,
  \label{eq:kirchhoff}
\end{equation}
where $a$ is the plate radius, $t$ its thickness, $\rho$ the density, $E$ Young's
modulus, $\nu$ Poisson's ratio, and $\lambda^{2}(n,s)$ a dimensionless coefficient
depending on the mode---$n$ nodal diameters, $s$ nodal circles---and on $\nu$. The
fundamental of a free disc is the $(2,0)$ mode: two nodal diameters, no nodal
circles, with a node at the centre, which allows the disc to be supported there
without appreciable damping.

Leissa~\cite{leissa1969} tabulates these coefficients in \S2.1.3, defining
$\lambda^{2}=\omega a^{2}\sqrt{\rho_{A}/D}$ with $a$ the \emph{radius}, $\omega$ the
circular frequency, $\rho_{A}$ the mass per unit \emph{area}, and
$D=Et^{3}/12(1-\nu^{2})$; substituting $\rho_{A}=\rho t$ and $\omega=2\pi f$ recovers
\eqref{eq:kirchhoff}. Table~2.5, from Colwell and Hardy~\cite{colwell1937} and
headed $\nu=0.33$ (its values are reproduced at $\nu=1/3$; Appendix~\ref{app:lambda}), gives $\lambda^{2}(2,0)=5.253$ and
$\lambda^{2}(3,0)=12.23$; Table~2.6, at $\nu=0.25$ and compiled by Leissa from earlier
work (his ref.~2.3), gives $5.513$ and $12.75$.

Rearranging \eqref{eq:kirchhoff} separates a geometry factor from a material factor:
\begin{equation}
  f \propto \frac{t}{a^{2}}\,c_{p},
  \qquad
  c_{p}=\sqrt{\frac{E}{\rho\left(1-\nu^{2}\right)}} .
  \label{eq:cp}
\end{equation}
Here $c_{p}$ is the extensional (plane-stress) plate wave speed, a material constant;
it is not the phase velocity of the flexural waves excited by the tap, which is
dispersive and frequency-dependent. $c_{p}$ enters \eqref{eq:kirchhoff} as the
material factor of the flexural resonance, and that is the sense in which it is
recovered here. Matching a precious metal's density does not in general imply
matching its $c_{p}$, which makes $c_{p}$ a useful independent material
discriminator; one scalar elastic parameter cannot, however, identify an alloy
uniquely. Recovering elastic constants from resonance is standardised as the impulse
excitation technique~\cite{astm1876}; what that standard assumes, and what a coin
denies, is a specimen of simple and accurately known geometry.

\subsection{Two obstacles}

\subsubsection{Thin-plate theory fails at coin proportions}
Kirchhoff theory neglects transverse shear and rotary inertia, which grow in
importance as thickness rises relative to radius; Mindlin's first-order
shear-deformation theory is the standard correction~\cite{mindlin1951}. A one-ounce
silver round has $a/t\approx6$, inside the region where the uncorrected theory
carries meaningful error. The correction is measured here rather than assumed, by
fitting
\begin{equation}
  f_{1}=C\,\frac{t_\mathrm{eff}/a^{2}}{1+\beta\left(t_\mathrm{eff}/a\right)^{2}} ,
  \label{eq:shear}
\end{equation}
with $\beta$ a dimensionless coefficient and $C$ a material-and-mode constant.
Comparison with \eqref{eq:kirchhoff} in the thin-plate limit gives
$C=\lambda^{2}(2,0)\,c_{p}/(2\pi\sqrt{12})$, so $C$ contains the elastic constants of
the calibration material, through $c_{p}$ and, via $\lambda^{2}$, through $\nu$.
Equation~\eqref{eq:shear} is a one-parameter rational approximation motivated by
Mindlin theory rather than a shear-correction factor in that theory's sense: $\beta$
absorbs transverse shear and rotary inertia into a single fitted constant, and is
reported as an empirical finite-thickness coefficient.

\subsubsection{A struck coin is not a uniform disc}
Relief removes material from the nominal cylinder, so caliper thickness overstates the
plate that actually vibrates, by an amount set by the design rather than the metal.
This is the obstacle that has kept acoustic coin testing empirical. We define a
\emph{fill factor}
\begin{equation}
  \varphi=\frac{V}{\pi a^{2} t},
  \qquad
  t_\mathrm{eff}=\varphi t = \frac{V}{\pi a^{2}} ,
  \label{eq:phi}
\end{equation}
where $V$ is the true volume obtained by liquid displacement. A machined disc has
$\varphi\approx1$; a struck coin has $\varphi<1$. Equation~\eqref{eq:phi} converts an
unmeasurable geometry into a measurable one and is the central contribution of this
work.

It supplies a \emph{volume-equivalent} thickness, not a dynamically equivalent one.
Local flexural rigidity varies as $h^{3}$, so two specimens of equal volume whose
material is distributed differently relative to the modal curvature will not resonate
identically. The fill factor converts an unmeasurable thickness field into a
measurable average; Section~\ref{sec:discussion}-A quantifies where that approximation
fails.

\subsection{The estimator and its sensitivities}
Inverting \eqref{eq:shear} with \eqref{eq:phi} gives the readout directly, in terms of
the three quantities actually measured:
\begin{equation}
\begin{split}
  c_{p,\mathrm{eff}}
   &=\frac{2\pi\sqrt{12}\,f_{1}a^{2}}{\lambda^{2}t_\mathrm{eff}}
    \left[1+\beta\!\left(\frac{t_\mathrm{eff}}{a}\right)^{\!2}\right]\\[2pt]
   &=\frac{2\pi^{2}\sqrt{12}\,f_{1}a^{4}}{\lambda^{2}V}
    \left[1+\beta\!\left(\frac{V}{\pi a^{3}}\right)^{\!2}\right].
\end{split}
  \label{eq:estimator}
\end{equation}
Caliper thickness cancels between $\varphi$ and $t$, and specimen mass does not enter:
the acoustic readout depends only on $f_{1}$, $a$, $V$, $\lambda^{2}$ and $\beta$.
Mass enters separately, through the Archimedes density $\rho=m/V$ on which the
decision rule of Section~\ref{sec:results}-G acts. Writing
$q=\beta\left(t_\mathrm{eff}/a\right)^{2}$, the logarithmic sensitivities are
\begin{equation}
  \frac{\partial\ln c_{p}}{\partial\ln f_{1}}=1,\quad
  \frac{\partial\ln c_{p}}{\partial\ln V}=\frac{q-1}{1+q},\quad
  \frac{\partial\ln c_{p}}{\partial\ln a}=4-\frac{6q}{1+q},
  \label{eq:sens}
\end{equation}
with $\partial\ln c_{p}/\partial\ln\lambda^{2}=-1$ and
$\partial\ln c_{p}/\partial\ln\beta=q/(1+q)$. Over the calibration discs $q$ runs from
$0.008$ to $0.134$; over the struck coins, where $V$ and $a$ actually enter, it is
$0.001$ to $0.022$, so displaced volume propagates at $-0.96$ to $-1.00$ and radius at
$+3.9$ to $+4.0$ rather than $+1$: radius appears both in $a^{2}$ and in the
volume-derived $t_\mathrm{eff}$. Section~\ref{sec:budget} shows
that this reweighting does not change the ordering of the error budget.

\subsection{Choice of \texorpdfstring{$\lambda^{2}$}{lambda squared}}
Because $\lambda^{2}$ depends on $\nu$ and Leissa tabulates the free plate at only two
Poisson's ratios, coefficients were obtained by solving the free-edge characteristic
equation directly---boundary conditions $M_{r}=0$ and Kirchhoff shear $V_{r}=0$,
deflection $A J_{n}(kr)+C I_{n}(kr)$, a $2\times2$ determinant, $\lambda^{2}=(ka)^{2}$.
The solver reproduces Leissa's Table~2.6 to within \SI{0.04}{\percent} and Table~2.5
to within \SI{0.09}{\percent} on the six low modes tabulated at both ratios. Values used here are
given in Table~\ref{tab:lambda}.

For a specimen of unknown material $\nu$ is itself unknown, so $\lambda^{2}$ cannot be
selected without presupposing part of the answer. The readout is therefore a
$\nu$-conditioned estimate of $c_{p}$; Section~\ref{sec:discussion}-D bounds the
resulting systematic at $\pm$\SI{3}{\percent} across $\nu=0.28$--$0.37$. This fixes
what \emph{absolute} means throughout this paper: computed from the measured resonance
and geometry through a physical model, rather than retrieved from a coin-specific
frequency library. It does not mean an assumption-free determination of the material's
intrinsic elastic constants.

\begin{table}[!t]
\caption{Free-plate coefficients used, computed from the characteristic equation}
\label{tab:lambda}
\centering
\begin{tabular}{lccc}
\toprule
Material & $\nu$ & $\lambda^{2}(2,0)$ & $\lambda^{2}(3,0)$\\
\midrule
Tungsten            & 0.28 & 5.421 & 12.564\\
Steel (304)         & 0.29 & 5.390 & 12.502\\
Aluminium           & 0.35 & 5.196 & 12.109\\
Copper, brass       & 0.34 & 5.229 & 12.177\\
Silver              & 0.37 & 5.128 & 11.969\\
\bottomrule
\end{tabular}
\end{table}

Using a single $\lambda^{2}=5.253$ across materials, as is common practice, introduces
a systematic error of up to \SI{2.4}{\percent}, largest on silver.

\subsection{Reference values}
Every reference wave speed quoted here is computed from \eqref{eq:cp} rather than
taken from a table, so the constants behind it are given in
Table~\ref{tab:handbook}~\cite{kayelaby,crc}. They are not single numbers. Tabulated
Young's moduli for the pure metals span several per cent---copper from
\SIrange{110}{128}{\giga\pascal}, silver from \SIrange{76}{83}{\giga\pascal}---so the
implied $c_{p}$ carries a half-width of \SIrange{0.7}{3.8}{\percent} before any measurement
is made. Agreement closer than that spread cannot be read as accuracy, and a comparison
against one chosen value inside the range is not a test.

This work therefore inverts the comparison. Rather than asking whether the recovered
$c_{p}$ matches a selected reference, it recovers the Young's modulus implied by the
measurement, $E=c_{p}^{2}\rho(1-\nu^{2})$ with the specimen's own measured density, and
asks whether that value falls inside the accepted range for the claimed material. This
is what ASTM~E1876 does with resonance data~\cite{astm1876}, it removes the dependence
on which end of a tabulated range is quoted, and it makes the test falsifiable in a way
a percentage agreement is not.

\begin{table}[!t]
\caption{Accepted elastic constants and the reference $c_{p}$ they imply. The modulus
is a range, not a value, and the wave-speed column inherits that spread. Gold and
tungsten are listed for the density-matching argument of Section~\ref{sec:theory} and
are not specimens here.}
\label{tab:handbook}
\centering
\small
\begin{tabular}{lcrcc}
\toprule
Material & $E$ (GPa) & $\rho$ & $\nu$ & $c_{p}$ (m\,s$^{-1}$)\\
 & accepted & (kg\,m$^{-3}$) & & implied\\
\midrule
Aluminium         & 68--70   & 2700  & 0.35 & 5357--5436\\
Copper            & 110--128 & 8960  & 0.34 & 3726--4019\\
Silver (fine)     & 76--83   & 10490 & 0.37 & 2897--3028\\
Silver (sterling) & 76--83   & 10360 & 0.37 & 2915--3047\\
Brass (60/40)     & 96--110  & 8500  & 0.34 & 3574--3825\\
Steel (plain C)   & 190--210 & 7850  & 0.29 & 5141--5404\\
\midrule
Gold              & 78       & 19300 & 0.44 & 2240\\
Tungsten          & 411      & 19250 & 0.28 & 4813\\
\bottomrule
\end{tabular}
\end{table}

\section{Method}\label{sec:method}

\subsection{Specimens}
Five machined aluminium discs were used for calibration, at two nominal thicknesses (3.18 and \SI{6.58}{\milli\meter})
across four nominal diameters (25.05, 34.00, 48.00 and \SI{68.00}{\milli\meter}),
giving $a/t$ from 2.58 to 10.75. Two discs share a diameter at differing thickness and
two share a thickness at differing diameter, so the geometry sweep is two-dimensional
rather than a single line.

The supplier's alloy designation and temper were not recorded, and this is stated rather
than guessed. The consequence is bounded and small. The fitted constant $C$ does
incorporate the aluminium's elastic properties, since
$C=\lambda^{2}(2,0)\,c_{p}/(2\pi\sqrt{12})$, but it is determined experimentally from
$f_{1}$ and the measured geometry and does not require prior knowledge of them; the
finite-thickness coefficient $\beta$ is identified from the variation of the residuals
across the five disc geometries. The alloy enters only when $C$ is converted to $c_{p}$,
which requires choosing $\lambda^{2}$, and therefore $\nu$. Taking the two conventions that bracket the common wrought alloys---$\nu=0.35$
with $\lambda^{2}=5.196$, and $\nu=0.33$ with $\lambda^{2}=5.253$---gives
$c_{p}=5413$ and \SI{5354}{\meter\per\second}, and a recovered modulus of
\SI{69.6}{\giga\pascal} and \SI{69.1}{\giga\pascal} respectively. Both lie inside the
\SIrange{68.9}{70.0}{\giga\pascal} spanned by commercially pure, 1100-series and
6061 aluminium, and the measured density, \SI{2707}{\kilo\gram\per\cubic\meter}, does
not discriminate between them either. The alloy question is therefore open but
immaterial at the precision of this work.

The reference set comprises seventeen specimens of stated composition: eight privately
minted 0.999 fine copper rounds, eight silver specimens (0.999 fine
rounds from two mints, a 2003 American Silver Eagle, and a machined sterling disc), and
a machined tungsten disc included to test the density criterion against the standard
high-density substitute. Six of the copper rounds (R1--R6) carry a displacement
measurement and form the copper validation set; the two five-ounce rounds (R7, R8)
have none and enter no result that requires one. All seventeen, and the five aluminium calibration discs, were
tested with a neodymium magnet and none is ferromagnetic. Stated compositions are mass
fractions: 0.999 fine denotes \SI{99.9}{\percent} of the named metal by mass, and
sterling \SI{92.5}{\percent} silver by mass.

Five additional bullion-like rounds of nominal one-ounce size, designated C1--C5,
complete the set; four are ferromagnetic. They were included to evaluate the method on
specimens of unverified composition and provenance. Their designs resemble established
bullion issues---two Australian designs (a crocodile and a bird), a United States
silver dollar, an Australian gold-coloured design and a Chinese issue. They were
obtained commercially and were selected because their authenticity was in doubt, but
neither their authenticity nor their elemental composition has been independently
established. Their measured density, magnetic response and acoustic wave speed are
compared with reference material properties in Section~\ref{sec:results}, which
describes them only as \emph{consistent with} candidate materials, never as
identified: the labels ``steel-like'' and ``brass-like'' used for them below describe
consistency with the measured properties and do not constitute material
identification.

\subsection{Metrology}
Thickness was measured with digital calipers of \SI{0.01}{\milli\meter} resolution at
three or four positions and averaged; diameter three times. The display resolution is
not the uncertainty in the radius the estimator needs. Diameters were recorded to
\SI{0.1}{\milli\meter} for the reference specimens and to \SI{0.5}{\milli\meter} for
the unverified rounds C1--C5, whose rims and edges were the least well defined. On a struck coin
the limiting quantity is where the edge is taken to be: a raised rim, edge reeding or
lettering, and small departures from circularity make the radius of the vibrating plate
uncertain well above the display resolution, and three readings at different
orientations sample that only partly. The
radius uncertainty carried into Section~\ref{sec:budget} is therefore the larger of
the recording interval and the scatter of the readings. Mass and displaced mass
were measured three times each, first on a balance of \SI{0.1}{\gram} resolution and
then on an analytical balance of \SI{0.1}{\milli\gram} resolution; the finer readings
are used wherever they exist. Dry mass was remeasured for every specimen except R6 and
aluminium disc~1, and displaced mass for every struck specimen except R6, R10 and R17;
R6 exceeds the capacity of the finer balance in the displacement configuration.
True volume was obtained by displacement: the
specimen was suspended on a thread and lowered into distilled water on a balance tared
to the filled vessel, so that a single reading gives the displaced mass directly. Three
such readings were taken per specimen and averaged. Water density was taken as
\SI{0.99878}{\gram\per\cubic\centi\meter} at the recorded \SI{17}{\celsius}.

\begin{figure}[!htbp]
\centering
\includegraphics[width=\columnwidth]{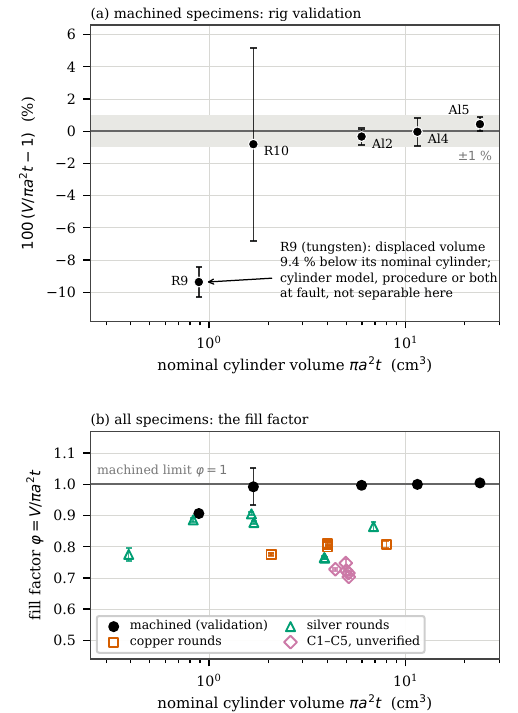}
\caption{(a) Rig validation. For a machined specimen the displaced volume must equal
the nominal cylinder, so the residual is the rig's error. Aluminium discs~4 and~5 fall
inside \SI{\pm1}{\percent}, disc~2 (at \SI{0.1}{\milli\gram}) to \SI{-0.33}{\percent};
the machined sterling disc is consistent with zero within one \SI{0.1}{\gram} division.
The tungsten disc is not: at \SI{-9.4}{\percent} its error bar excludes parity by a wide
margin. A mismatch of that size indicates that the ideal-cylinder assumption, the
measurement procedure, or both fail for this specimen; without independent dimensional
mapping or a composition measurement their contributions cannot be separated. Its
Archimedes density, \SI{19487}{\kilo\gram\per\cubic\meter}, is \SI{1.2}{\percent} above
the handbook value for pure tungsten---compatible with a small displacement bias, but
not by itself proof of composition. Its $\varphi=0.906$ should not be read as relief.
(b) The same comparison for every
specimen, expressed as the fill factor. Struck coins fall to 0.70--0.91; the machined
specimens mark the $\varphi=1$ limit. Error bars are the displaced-mass uncertainty of
Section~\ref{sec:method}-B: three readings and a \SI{0.5}{\percent} floor at
\SI{0.1}{\milli\gram}, one division for the three specimens measured only at
\SI{0.1}{\gram}.}
\label{fig:volume}
\end{figure}

The displacement rig was validated against machined discs of independently known
geometry (Table~\ref{tab:archimedes}, Fig.~\ref{fig:volume}(a)), for which the displaced
volume must by construction equal the nominal cylinder. In volume, discs~4 and~5 agree
to \SI{-0.04}{\percent} and \SI{+0.43}{\percent} at \SI{0.1}{\gram} resolution, and
disc~2 to \SI{-0.33}{\percent} at \SI{0.1}{\milli\gram}. A second check
is available in the fine-silver rounds, which at the finer resolution all read
\SIrange{10530}{10579}{\kilo\gram\per\cubic\meter}, a mean of $10559\pm21$ against a
handbook \num{10490}: \SI{0.65}{\percent} high, the same sign as the disc~2
discrepancy. Both point to a small systematic in displaced volume that repeating a
reading does not sample. Candidate mechanisms act in both directions---thread buoyancy
and a trapped bubble add displaced volume and lower the inferred density, while wetting
forces on the suspension thread reduce the balance reading and raise it---so the error
need not have the same sign on every specimen; disc~5 reads \SI{0.43}{\percent} low in
density.
The displaced-mass uncertainty at the finer resolution is therefore taken as the
standard deviation of the three readings combined in quadrature with a
\SI{0.5}{\percent} systematic floor. That floor is a judgement, not a calibration: it
is set between the two available indications, \SI{0.34}{\percent} on disc~2 and
\SI{0.65}{\percent} on the fine-silver mean, and rests on one machined disc and one
material. For R6, R10 and R17, which were not remeasured, the uncertainty remains one
\SI{0.1}{\gram} division.

The fine-silver excess may also be a bias rather than scatter. Silver of 0.999 fineness
cannot be denser than pure silver, so a \SI{0.65}{\percent} excess indicates that
displaced volume was under-read on those specimens. If the same under-read applied to
every displacement measurement, correcting for it would lower each density by
\SI{0.65}{\percent}, each recovered $c_{p}$ by about \SI{0.6}{\percent}, and each
recovered modulus by about \SI{2}{\percent}: copper from 119.1 to
\SI{116.8}{\giga\pascal}, still inside its accepted range; fine silver from 75.1 to
\SI{73.7}{\giga\pascal} and the steel-like rounds C1--C4 from 189.4 to
\SI{185.9}{\giga\pascal}, moving both from the edge of their accepted ranges to below
them. No density decision and no group ordering changes. The correction is not applied,
because the machined disc measured on the same balance reads high by only half as much
and the offset has not been calibrated independently; the recovered moduli of
Table~\ref{tab:bands} should be read with this possible common offset in mind.

\begin{table}[!t]
\caption{Validation of the displacement rig against machined discs}
\label{tab:archimedes}
\centering
\begin{tabular}{lccccc}
\toprule
Disc & $m$ (\si{\gram}) & $m_\mathrm{disp}$ (\si{\gram})
     & $\rho_\mathrm{geom}$ & $\rho_\mathrm{Arch}$ & Diff.\\
     &                  &
     & \multicolumn{2}{c}{(\si{\kilo\gram\per\cubic\meter})} & \\
\midrule
2$^{*}$ & 16.1845 & 5.9480 & 2708.6 & 2717.7 & \SI{+0.336}{\percent}\\
4 & 31.100 & 11.467 & 2707.8 & 2708.9 & \SI{+0.039}{\percent}\\
5 & 64.733 & 23.967 & 2709.4 & 2697.7 & \SI{-0.433}{\percent}\\
\midrule
\multicolumn{6}{@{}l}{\footnotesize $^{*}$\SI{0.1}{\milli\gram} balance; discs 4 and 5 at \SI{0.1}{\gram}}\\
\bottomrule
\end{tabular}
\end{table}

\subsection{Uncertainty budget}\label{sec:budget}
Table~\ref{tab:budget} collects the contributions, each weighted by its sensitivity
coefficient from \eqref{eq:sens}. The ordering is the useful result. At
\SI{0.1}{\gram} resolution the displaced-volume term exceeded the frequency term by one
to two orders of magnitude on every specimen, and by a factor of two hundred on the
smallest. At \SI{0.1}{\milli\gram} it falls to \SIrange{0.5}{2.6}{\percent}, set mainly
by the \SI{0.5}{\percent} systematic floor rather than by resolution: still the largest
\emph{instrumental} term for the reference specimens, but now comparable to or smaller
than the uncertainty in the reference values themselves and in $\lambda^{2}$ when $\nu$
is unknown. Radius, which the naive treatment would place last, enters at a sensitivity
of four once its appearance inside $t_\mathrm{eff}$ is accounted for. For the reference
specimens, whose diameters were read to \SI{0.1}{\milli\meter}, it contributes
\SIrange{0.2}{0.6}{\percent}, below volume; for the unverified rounds C1--C5, whose diameters were
read to \SI{0.5}{\milli\meter}, it contributes \SIrange{1.4}{2.8}{\percent} and is the
largest term. The uncertainty $u$ of Table~\ref{tab:specimens} combines the two.

That $u$ is a measurement-derived uncertainty, not a total predictive one. It omits the
Poisson's-ratio term for a specimen of unknown material (last row of
Table~\ref{tab:budget}), model error from non-uniform relief
(Section~\ref{sec:discussion}-A), any systematic bias in the displacement measurement
beyond the \SI{0.5}{\percent} floor (Section~\ref{sec:method}-B), and the error in
transferring $\beta$ from aluminium to other materials (Section~\ref{sec:results}-B).
A \SI{0.6}{\percent} error bar therefore does not mean that the recovered material
property is accurate to \SI{0.6}{\percent}.

\begin{table}[!t]
\caption{Contributions to the uncertainty in $c_{p,\mathrm{eff}}$, with the
sensitivity coefficient $\partial\ln c_{p}/\partial\ln x$ of \eqref{eq:sens}
evaluated over the struck coins ($q=0.001$--$0.022$). Only the $V$ and $a$ rows enter
the per-specimen $u$ of Table~\ref{tab:specimens} and Figs.~\ref{fig:bands}--\ref{fig:parity};
the remaining rows, and model error from non-uniform relief, are not included in $u$.}
\label{tab:budget}
\centering
\small
\begin{tabular}{llcr}
\toprule
Quantity & Source & Sens. & Contrib.\\
\midrule
$V$ (\SI{0.1}{\milli\gram}) & repeat${}+{}$floor & $-0.96$ to $-1$ & 0.5--2.6\,\%\\
$V$ (\SI{0.1}{\gram})       & division        & $-0.96$ to $-1$ & 1.6--6\,\%\\
$a$, reference      & \SI{0.1}{\milli\meter} read & $+3.9$ to $+4$ & 0.2--0.6\,\%\\
$a$, C1--C5         & \SI{0.5}{\milli\meter} read & $+3.9$          & 1.4--2.8\,\%\\
Frequency $f_{1}$   & session remount  & $+1$               & $0.15$\,\%\\
Water density       & $\pm$\SI{1}{\celsius} & $+1$               & $<0.02$\,\%\\
$\beta$             & calibration fit  & $q/(1+q)$          & $<0.63$\,\%\\
$\lambda^{2}$, $\nu$ known   & interpolated & $-1$               & $<0.1$\,\%\\
$\lambda^{2}$, $\nu$ unknown & 0.28--0.37 & $-1$               & $\pm3$\,\%\\
\bottomrule
\end{tabular}
\end{table}

Caliper thickness and specimen mass do not appear, because neither enters
\eqref{eq:estimator}. Both enter the Archimedes density instead, at
$\partial\ln\rho/\partial\ln m=+1$ and $\partial\ln\rho/\partial\ln V=-1$, contributing
below \SI{0.01}{\percent} and \SIrange{0.5}{2.6}{\percent} respectively to the
quantity the decision rule acts on at the finer resolution.

\subsection{Acoustic apparatus}
Measurements were made inside an enclosure of pyramid acoustic foam on a plastic box
frame: $30\times30$\,\si{\centi\meter} panels on the left and right walls,
a $15\times30$\,\si{\centi\meter} panel at the back, cloth over top and
bottom, front open for striker access. The open front admits some room noise but
permits manual excitation and removes a hard reflector from immediately in front of the
microphone.

Specimens were supported at the centre---a node of the $(2,0)$ mode---on a 3D-printed
PLA fixture with two pegs carrying silicone contact tips. Repeatable centring was
achieved by marking each specimen's centre and aligning it to the fixture. Audio was
captured with a Tascam DR-07X at \SI{96}{\kilo\hertz} and 24 bits in two channels, of
which the analysis uses channel~0 (left) only, all
automatic level processing disabled, microphone \SIrange{2}{3}{\centi\meter} from the
rim. Excitation was manual throughout: a \SI{31}{\gram} metal striker with a
\SI{3}{\milli\meter} flat circular tip, applied at the rim, an antinode of the mode of
interest.

\subsection{Analysis pipeline}
Recordings were processed by a single Python program (Python 3.13.5, NumPy 2.1.3,
SciPy 1.15.3, soundfile 0.14.0) emitting one row per detected tap. Onsets are declared
where the Hilbert envelope, smoothed over \SI{3}{\milli\second}, exceeds
\SI{15}{\percent} of the file maximum, with \SI{1.0}{\second} minimum spacing. The
analysis segment begins \SI{6}{\milli\second} after onset and runs \SI{0.35}{\second},
or \num{33600} samples. It is Hann-windowed and transformed on a grid whose length is
the next power of two at or above four times the sample count,
$2^{18}=\num{262144}$ points, giving \SI{0.366}{\hertz} bins at \SI{96}{\kilo\hertz}.
Relative to the \num{33600}-sample analysis window the net zero-padding factor is
therefore \num{7.8}, not \num{4}: the fourfold multiplication is applied to the raw
sample count and the result is then rounded up to a radix-2 length. Zero-padding
interpolates the spectrum and does not improve true spectral resolution, which for a
\SI{0.35}{\second} segment is \SI{2.9}{\hertz}; peak positions are refined further by
parabolic interpolation on the log-magnitude spectrum, and the measured
strike-to-strike scatter of \SIrange{0.008}{0.084}{\percent} rather than the bin width
is the operative bound on peak-position precision.

\begin{figure}[!htbp]
\centering
\includegraphics[width=\columnwidth]{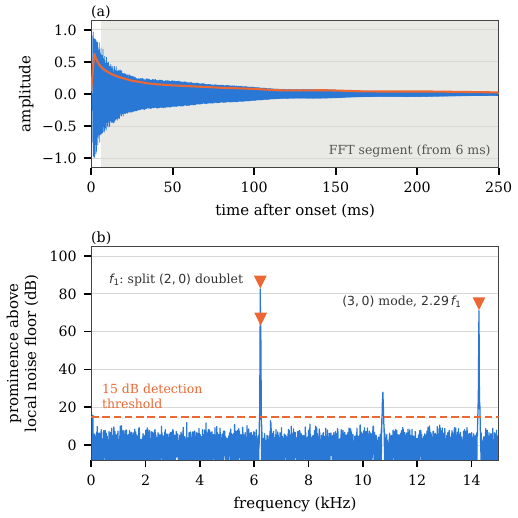}
\caption{One tap on a one-ounce copper round, as the pipeline sees it. (a) The
recorded waveform and its Hilbert envelope, with the \SI{0.35}{\second} analysis
segment shaded. (b) The windowed spectrum expressed as prominence above a
\SI{400}{\hertz} running-median noise floor; markers show the peaks that clear both
the \SI{15}{\deci\bel} threshold and the \SI{3}{\percent} magnitude criterion. The
two markers at $f_{1}$ are the resolved members of the split $(2,0)$ doublet; the
peak at $2.29\,f_{1}$ is the $(3,0)$ mode, not a harmonic.}
\label{fig:spectrum}
\end{figure}

Signal-to-noise is defined relative to a local floor---the log spectrum minus a
\SI{400}{\hertz} running median of itself. Peaks require \SI{15}{\deci\bel} of local
prominence and \SI{8}{\hertz} separation, above a \SI{1000}{\hertz} floor that excludes
enclosure air resonances. The lowest peak with magnitude at least \SI{3}{\percent} of
the strongest is taken as $f_{1}$. The \SI{1000}{\hertz} floor is a deliberate choice
with one known cost, discussed in Section~\ref{sec:results}-F: a specimen whose true
fundamental lies below it has that fundamental replaced by its first detected overtone.

Ring-down time is fitted after band-passing to $f_{1}\pm\SI{20}{\hertz}$ with a
zero-phase fourth-order Butterworth filter. This step is necessary rather than
cosmetic: when the $(2,0)$ mode splits, the two members beat and the broadband envelope
is $e^{-t/\tau}\lvert\cos(\pi\Delta f t)\rvert$, which repeatedly approaches zero, so a
single-exponential fit to it is meaningless. The fit is an unweighted linear regression
of the log envelope against time, running from the envelope maximum
plus \SI{5}{\milli\second} to \SI{20}{\deci\bel} below, capped at
\SI{0.8}{\second} and truncated at the next tap's onset so that no fit borrows audio
from the following ring. Fits spanning less than \SI{20}{\milli\second}, or returning a
non-decaying slope, are discarded, and only fits with $R^{2}\ge0.90$ are reported.
Quality factor is $Q=\pi f_{1}\tau$.

This prominence statistic has a floor in $Q$ as well as in amplitude, which bounds the
class of specimen the method can address at all. Several properties of the analysis
interact to produce it. A mode's linewidth is approximately $f/Q$, and a broadened line
raises the running-median baseline it must stand clear of; but a linewidth of
\SI{400}{\hertz}, the width of the median window, is reached only when
$Q\approx f/\SI{400}{\hertz}$---about 15 at \SI{6}{\kilo\hertz} and 50 at
\SI{20}{\kilo\hertz}---so linewidth alone does not set the boundary observed. A low-$Q$
mode also decays quickly, with $\tau=Q/\pi f$, about \SI{3}{\milli\second} for $Q=200$
at \SI{20}{\kilo\hertz}: comparable to the \SI{6}{\milli\second} delay before the
analysis segment begins, and short against a Hann-windowed segment whose weight is near
zero at its start, so little of the ring's energy reaches the spectrum. Together with
the fixed prominence threshold these effects set a boundary that depends on frequency,
and it was established empirically rather than derived. Synthetic decaying sinusoids at
\SIrange{16}{24}{\kilo\hertz} were injected into 24 segments of recorded inter-tap
noise and analysed with the pipeline's prominence statistic on a \SI{75}{\milli\second}
segment beginning \SI{6}{\milli\second} after onset. No amplitude up to half of full
scale produced a detection at $Q\le100$ (16--\SI{20}{\kilo\hertz}) or $Q\le150$
(22--\SI{24}{\kilo\hertz}), while every frequency was detected in all realizations by
$Q\approx220$. Every specimen
reported here is far above that threshold---aluminium 872--9909, copper 2982--3269,
silver 1456--7013---so it does not affect any result in this work, but it does mean a
sufficiently damped specimen is invisible to the acoustic axis regardless of how hard it
is struck.

Taps are rejected if $f_{1}$ carries less than \SI{25}{\deci\bel} of prominence, or if
$f_{1}$ lies more than \SI{2}{\percent} from the median of the same recording. Of 2163
taps in the production run, 1937 (\SI{89.6}{\percent}) were accepted. Specimen-level
$f_{1}$ is the mean of the accepted taps. For the five aluminium calibration discs it is
instead the mean of the dominant cluster, where clusters are split at consecutive gaps
exceeding \SI{0.1}{\percent} (Appendix~\ref{app:provenance}); applying that rule to the
other specimens would change no fundamental by more than \SI{0.11}{\percent}, below the
displaced-volume term.

\section{Results}\label{sec:results}

\subsection{Repeatability}
Strike-to-strike scatter in $f_{1}$ within a take has median \SI{0.050}{\percent} over
96 coin takes (interquartile range \SIrange{0.008}{0.084}{\percent}). Between-session
reproducibility is the binding figure: four aluminium discs remounted and re-measured
on a separate day shifted by $-0.026$, $-0.133$, $-0.159$ and
\SI{-0.222}{\percent}---all in the same direction, RMS \SI{0.152}{\percent}. A stronger
statement follows from fitting the calibration constant independently from each session
at fixed $\beta$. Restricted to the four discs measured in both, the two sessions give
$1290.50\pm2.41$ and $1289.66\pm2.79$\,\si{\hertz\meter}, agreeing to
\SI{0.07}{\percent}---from independent mountings on different days, with the same
estimator and the same parameters.

\subsection{Calibration on aluminium}
\begin{figure}[!htbp]
\centering
\includegraphics[width=\columnwidth]{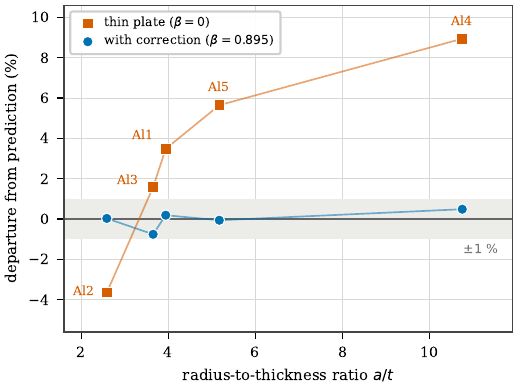}
\caption{Departure of measured $f_{1}$ from prediction against radius-to-thickness
ratio, before and after the finite-thickness correction of \eqref{eq:shear}. The
uncorrected residuals are structured---monotonic in $a/t$ with a single crossover---and
span 12.6 percentage points; one fitted parameter reduces the worst case to
\SI{0.75}{\percent} and removes the structure. Both curves are computed live from the
five discs' measured geometry and fundamentals.}
\label{fig:residuals}
\end{figure}

Fitting the uncorrected thin-plate prediction $f_{1}=C(t/a^{2})$ through the origin
gives $C=\SI{1182.8}{\hertz\meter}$ with $R^{2}=0.99530$. That would ordinarily be
reported as confirmation. It is not: ordered by increasing $t/a$ the residuals run
$+8.9$, $+5.6$, $+3.5$, $+1.6$ and \SI{-3.6}{\percent}, thin discs ringing high and
thick discs low, monotonically with a single crossover
(Fig.~\ref{fig:residuals}). A structured residual spanning 12.6 percentage points, on a
measurement good to \SI{0.15}{\percent}, indicates a missing physical term.

Fitting \eqref{eq:shear} jointly, unweighted over the five discs, gives
$\beta=0.895\pm0.047$, $C=1292.2\pm6.1$\,\si{\hertz\meter}, $R^{2}=0.9999672$, RMS
residual \SI{0.41}{\percent} and worst case \SI{0.75}{\percent}. The quoted
uncertainties are standard errors from a two-parameter fit to five points, three
degrees of freedom; $\beta$ is an empirical coefficient for these discs and this range
of aspect ratio, not a universal or independently validated physical constant. Residuals throughout
are quoted on $f_{1}$ relative to the fitted prediction; the deviation of a specimen's
recovered $c_{p}$ from the weighted mean across discs is a different quantity and is
larger. Weighting each disc by
its own strike-to-strike scatter instead gives $\beta=0.913\pm0.052$, within the
unweighted uncertainty; the unweighted fit is reported because it does not let the two
best-behaved discs dominate a five-point fit.

The calibration does not depend on which revision of the per-tap analysis supplies
$f_{1}$. Refitting from each of the two independent output files that cover this
session returns $\beta=0.895$ and $0.897$ and $C=1292.2$ and
\SI{1292.6}{\hertz\meter}---a spread in $\beta$ of one-twentieth of its own uncertainty
and in $C$ of \SI{0.03}{\percent}---with worst-case residuals of \SI{0.75}{\percent} and
\SI{0.83}{\percent}. The largest single input difference between the two files is
disc~4 at \SI{0.36}{\percent} in $f_{1}$, and disc~4 is the extreme
aspect-ratio point that anchors the thin end of the fit.

$\beta$ is fitted on aluminium and applied to every material. The first-order
finite-thickness coefficient scales approximately as $2(1+\nu)/\kappa$, so transferring
it across $\nu=0.28$--$0.37$ perturbs $\beta$ by at most about \SI{4}{\percent} and, at
coin proportions, $c_{p}$ by less than \SI{0.1}{\percent}, well below the volume term.

With $\lambda^{2}(2,0)=5.196$ and the measured geometric density
\SI{2706.6}{\kilo\gram\per\cubic\meter}, the calibration returns
$c_{p}=5412\pm25$\,\si{\meter\per\second} (per-disc mean and standard deviation,
\SI{0.46}{\percent}) against a handbook \SI{5436}{\meter\per\second}, a deviation of
\SI{-0.44}{\percent}. The fitted constant $C$, through
$C=\lambda^{2}(2,0)\,c_{p}/(2\pi\sqrt{12})$, implies \SI{5413}{\meter\per\second},
agreeing with the per-disc mean to \SI{0.02}{\percent}.
Taking $E=c_{p}^{2}\rho(1-\nu^{2})$ from the per-disc mean gives
$E=\SI{69.56}{\giga\pascal}$ against \SI{70}{\giga\pascal}. Poisson's ratio for pure
aluminium is quoted between 0.33 and 0.35; the lower value would shift $c_{p}$ by
\SI{-1.1}{\percent}, a stated systematic.

\subsection{Material bands}
\begin{figure}[!htbp]
\centering
\includegraphics[width=\columnwidth]{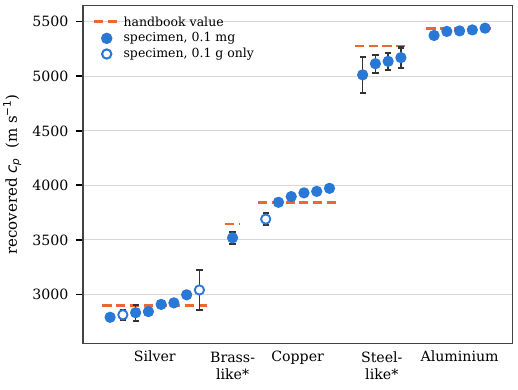}
\caption{Recovered plate wave speed by specimen, with per-specimen uncertainty $u$ of
Table~\ref{tab:specimens}, grouped by stated or (asterisked rounds of unverified composition) inferred
material. Hollow markers are the specimens measured only at
\SI{0.1}{\gram}. Dashed lines mark handbook values. The copper band mean, no member of
which entered the calibration, lands within \SI{0.9}{\percent} of handbook, with
individual specimens scattering by \SI{2.6}{\percent} about it. The groups are observed
groupings of point estimates, not uncertainty-resolved composition classes; aluminium
is the calibration material. Error bars are measurement-derived (displaced volume and
radius) and omit the Poisson's-ratio, relief-model, displacement-bias and
$\beta$-transfer terms of Section~\ref{sec:budget}.}
\label{fig:bands}
\end{figure}

Applying $t_\mathrm{eff}=\varphi t$ with per-material $\lambda^{2}$ produces the bands
of Table~\ref{tab:bands} and Fig.~\ref{fig:bands}. Fill factors span
0.70--0.99: the machined sterling disc reads 0.99 as it must, the most
deeply relieved rounds 0.70.

\begin{table}[!t]
\caption{Recovered plate wave speed by material, and the Young's modulus it implies
through $E=c_{p}^{2}\rho(1-\nu^{2})$ with each specimen's own measured density. The test
is whether the recovered modulus falls inside the accepted range, not whether $c_{p}$
matches a chosen value. The two groups of unverified rounds are labelled by inference from density,
magnetism and wave speed, so their comparison is conditional on that assignment.
``Edge'' marks a group mean below the accepted range by less than its standard error.
The groups are observed groupings of point estimates; they are not shown to be
resolved from one another within the measurement uncertainties.}
\label{tab:bands}
\centering
\small
\begin{tabular}{@{}lcrccl@{}}
\toprule
Material & $n$ & $c_{p}$ (SD) & $E$ rec. & $E$ acc. & \\
 & & (m\,s$^{-1}$) & (GPa) & (GPa) & \\
\midrule
Silver, fine      & 7 & 2872 (73)  & $75.1(40)$   & 76--83   & edge\\
Silver, sterling  & 1 & 3040 & $80.9$             & 76--83   & \textbf{in}\\
Brass-like$^{\dagger}$  & 1 & 3518 & $91.3$           & 96--110  & low\\
Copper            & 6 & 3879 (103) & $119.1(67)$  & 110--128 & \textbf{in}\\
Steel-like$^{\dagger}$ & 4 & 5108 (68)  & $189.4(64)$  & 190--210 & edge\\
Aluminium$^{\ddagger}$ & 5 & 5412 (25)  & $69.6(6)$    & 68--70   & \textbf{in}\\
\bottomrule
\multicolumn{6}{@{}l}{\footnotesize $^{\dagger}$round of unverified composition; class inferred, not assayed}\\
\multicolumn{6}{@{}l}{\footnotesize $^{\ddagger}$calibration material; not an independent test}
\end{tabular}
\end{table}

The point estimates of the six groups do not overlap: the highest fine-silver reading,
\SI{2996}{\meter\per\second}, and the sterling disc at \SI{3040}{\meter\per\second}
both sit below the lowest copper reading, \SI{3690}{\meter\per\second}. That is an
observation about point estimates, not an uncertainty-based discrimination of six
composition classes. The sterling disc's \SI{6.1}{\percent} uncertainty, about
$\pm$\SI{185}{\meter\per\second}, overlaps the fine-silver readings, and the accepted
elastic constants of fine and sterling silver overlap in Table~\ref{tab:handbook}, so
the two silver groups are not resolved from each other. The brass-like and steel-like
groups are rounds of unverified composition, the first a single specimen, so
neither characterises a population; and aluminium, the calibration material, cannot
test the method at all. The copper band carries the weight of the argument. Six struck rounds spanning \SIrange{32.6}{47.0}{\milli\meter} and
$\varphi=0.78$--$0.81$, none of which entered the calibration, recover
$E=119.1\pm6.7$\,\si{\giga\pascal} against an accepted
\SIrange{110}{128}{\giga\pascal}. The mean sits at the centre of that range and the
scatter spans most of it: what is demonstrated is that the ensemble is consistent with
copper, not that a single measurement is accurate to better than a few per cent. The
lowest copper reading, R6, is one of the three specimens measured only at
\SI{0.1}{\gram}. Aluminium
cannot supply even that, being the material on which $C$ was fitted; copper is the
independent test, and the fill-factor correction passes it.

At \SI{0.1}{\gram} resolution the groups split cleanly into three inside their
accepted ranges and three below, by \SIrange{8}{13}{\percent}, and the three below were
the deeply relieved specimens. The finer balance removes most of that pattern. Fine
silver and the steel-like rounds now sit at the lower edge of their ranges, by
\SI{1.2}{\percent} and \SI{0.3}{\percent}, each within its standard error, and only
the single brass-like round remains clearly low, at \SI{-4.9}{\percent} against
generic brass. Most of the earlier deficit was therefore volume resolution, not relief; what
survives of the design effect is examined in Section~\ref{sec:discussion}-A.

\begin{figure}[!htbp]
\centering
\includegraphics[width=\columnwidth]{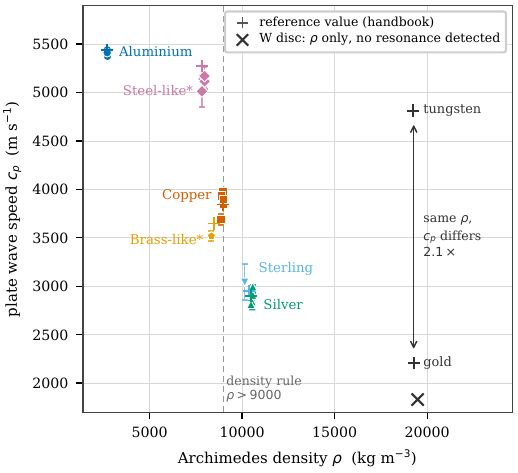}
\caption{Archimedes density against recovered plate wave speed, by specimen, with
per-specimen uncertainty $u$ of Table~\ref{tab:specimens}. Crosses mark
reference values; the dashed line is the decision rule's density threshold. Tungsten
and gold are shown as reference points only. The tungsten disc carries a density but no
acoustic reading---no resonance was detected from it---and is plotted at
the axis foot to mark the one specimen the density rule admits and the acoustic axis is
predicted, from handbook constants for pure tungsten, to reject; that prediction has
not been tested acoustically. Asterisked groups are rounds of unverified
composition. Error bars are measurement-derived only (Section~\ref{sec:budget}).}
\label{fig:rhocp}
\end{figure}

Figure~\ref{fig:rhocp} plots the two measured quantities against each other and shows
why the pair is stronger than either alone. Density alone admits everything to the
right of the decision threshold---fine silver, sterling and the tungsten disc
together---while the silver specimens' recovered $c_{p}$
(\SIrange{2790}{3040}{\meter\per\second}) lies about \SI{60}{\percent} below the
\SI{4810}{\meter\per\second} that handbook constants predict for pure tungsten. That
separation is predicted, not measured: no acoustic reading was obtained from either
tungsten specimen (Section~\ref{sec:results}-G). The reference values for gold
and tungsten, plotted for context, make the general point: the two differ by
\SI{0.3}{\percent} in density and by a factor of $2.1$ in $c_{p}$, so a density-matched
substitution is an elasticity mismatch by construction.

\begin{figure}[!htbp]
\centering
\includegraphics[width=\columnwidth]{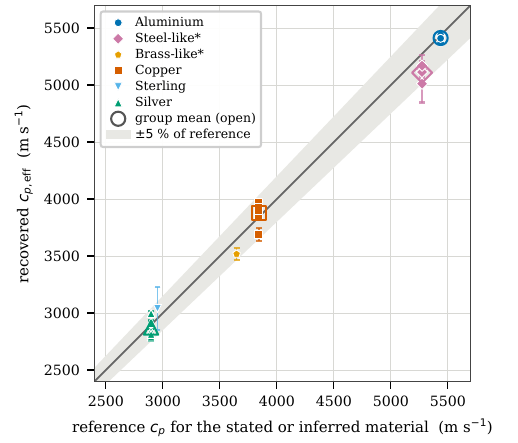}
\caption{Recovered $c_{p,\mathrm{eff}}$ against the reference value for each
specimen's stated material or, for the asterisked rounds of unverified composition, inferred material,
with the parity line and a $\pm$\SI{5}{\percent} band. Open
markers are group means, shown for groups of more than one specimen. Every specimen lies within about \SI{5}{\percent} of its
reference. Fine silver sits a few per cent low, consistent with the residual design
effect of Section~\ref{sec:discussion}-A; the brass-like and steel-like rounds also
sit low, but because their composition is inferred, that deficit cannot be separated
from a compositional difference from the reference alloy. Plotting
specimens rather than material averages shows the within-material scatter that a band
plot hides. Error bars are measurement-derived only (Section~\ref{sec:budget}).}
\label{fig:parity}
\end{figure}

\subsection{Complete specimen results}
Table~\ref{tab:specimens} lists every specimen with its measured geometry, fill
factor, density, fundamental, damping and recovered wave speed. The acoustic
quantities---$f_{1}$, $\tau$, $Q$ and $c_{p}$---are absent for the tungsten disc,
from which no resonance was detected; its geometry and density are
present, and the density is what the decision rule of Section~\ref{sec:results}-G
acts on. The two five-ounce copper rounds, R7 and R8, lack a displacement measurement
and therefore a fill factor: one has recordings and one has none. Neither belongs to
the six-round copper validation set (R1--R6), and neither enters any group mean,
recovered modulus or decision-rule result.

\begin{table*}[!t]
\caption{Complete specimen inventory and results. For struck specimens $u$ is the
relative uncertainty in $c_{p,\mathrm{eff}}$ from displaced mass and radius combined:
displaced mass as three readings with a \SI{0.5}{\percent} floor at
\SI{0.1}{\milli\gram}, or one \SI{0.1}{\gram} division for R6, R10 and R17; radius as
the standard error of three caliper readings or the reading resolution, whichever is
larger. Mass and density are from the finer balance wherever it was used (all but R6
and Al1). For the machined aluminium discs, whose volume is taken geometrically
($\varphi\equiv1$, confirmed by displacement on discs~2, 4 and~5 to within
\SI{0.5}{\percent}), $u$ is propagated from the caliper readings instead. All $c_{p}$
values use $\beta=0.895$. The $u$ values are measurement-derived, not total predictive
uncertainties: they exclude Poisson's-ratio uncertainty for a specimen of unknown
material, relief model error, displacement bias beyond the \SI{0.5}{\percent} floor and
$\beta$-transfer error (Section~\ref{sec:budget}). R7 and R8, the five-ounce copper
rounds, have no displacement measurement and are not part of the copper validation set
(R1--R6). For C1--C5 the material column gives the class the measurements are
consistent with, not an assayed composition.}
\label{tab:specimens}
\centering
\small
\begin{tabular}{llrrrrrrrrrrr}
\toprule
ID & Material & $d$ & $t$ & $m$ & $\varphi$ & $\rho$ & $f_{1}$ & SD & $\tau$ & $Q$ & $c_{p}$ & $u$\\
   &          & (mm) & (mm) & (g) &          & (kg\,m$^{-3}$) & (Hz) & (\%) & (ms) & & (m\,s$^{-1}$) & (\%)\\
\midrule
Al1 & Al & 25.05 & 3.181 & 4.20 & 1.000 & 2679 & 24821.2 & 0.020 & 17 & 1348 & 5424 & 0.1 \\
Al2 & Al & 34.00 & 6.581 & 16.18 & 1.000 & 2709 & 25953.9 & 0.029 & 50 & 4074 & 5415 & 0.1 \\
Al3 & Al & 48.00 & 6.581 & 32.45 & 1.000 & 2725 & 13728.9 & 0.054 & 44 & 1912 & 5373 & 0.1 \\
Al4 & Al & 68.00 & 3.163 & 31.14 & 1.000 & 2712 & 3525.1 & 0.066 & 895 & 9909 & 5440 & 0.1 \\
Al5 & Al & 68.00 & 6.579 & 64.73 & 1.000 & 2709 & 7111.3 & 0.038 & 39 & 872 & 5410 & 0.1 \\
R1 & Cu 0.999 & 32.60 & 2.480 & 14.42 & 0.776 & 8974 & 6779.7 & 0.091 & 151 & 3214 & 3944 & 0.61 \\
R2 & Cu 0.999 & 32.60 & 2.480 & 14.37 & 0.775 & 8964 & 6815.4 & 0.071 & 142 & 3037 & 3972 & 0.62 \\
R3 & Cu 0.999 & 39.30 & 3.310 & 28.84 & 0.799 & 8987 & 6225.2 & 0.043 & 152 & 2982 & 3844 & 0.57 \\
R4 & Cu 0.999 & 39.30 & 3.310 & 28.88 & 0.810 & 8878 & 6450.5 & 0.069 & 150 & 3035 & 3930 & 0.59 \\
R5 & Cu 0.999 & 39.30 & 3.310 & 29.07 & 0.806 & 8983 & 6360.7 & 0.097 & 155 & 3106 & 3896 & 0.56 \\
R6 & Cu 0.999 & 47.00 & 4.600 & 57.00 & 0.807 & 8849 & 5829.7 & 0.279 & 179 & 3269 & 3690 & 1.5 \\
R7 & Cu 0.999 & 64.00 & 6.070 & 143.90 & --- & --- & 4366.3 & 0.002 & 946 & 12974 & --- & --- \\
R8 & Cu 0.999 & 64.00 & 6.070 & 144.67 & --- & --- & --- & --- & --- & --- & --- & --- \\
R9 & W (machined) & 25.30 & 1.767 & 15.69 & 0.906 & 19487 & --- & --- & --- & --- & --- & --- \\
R10 & Ag sterling & 51.17 & 0.818 & 16.92 & 0.992 & 10140 & 887.1 & 0.368 & 687 & 1924 & 3040 & 6.1 \\
R11 & Ag 0.999 & 19.20 & 1.350 & 3.19 & 0.775 & 10530 & 7499.1 & 0.005 & 62 & 1456 & 2832 & 2.6 \\
R12 & Ag 0.999 & 26.57 & 1.500 & 7.80 & 0.886 & 10579 & 5138.2 & 0.138 & 123 & 1985 & 2922 & 0.69 \\
R13 & Ag 0.999 & 32.60 & 1.970 & 15.71 & 0.905 & 10553 & 4364.4 & 0.103 & 156 & 2139 & 2790 & 0.62 \\
R14 & Ag 0.999 & 39.30 & 3.200 & 31.23 & 0.763 & 10539 & 4178.6 & 0.095 & 158 & 2078 & 2842 & 0.62 \\
R15 & Ag 0.999 & 32.50 & 2.040 & 15.70 & 0.877 & 10575 & 4733.5 & 0.339 & 471 & 7013 & 2996 & 0.63 \\
R16 & Ag 0.999 & 40.60 & 2.980 & 31.21 & 0.765 & 10575 & 3748.3 & 0.124 & 524 & 6170 & 2908 & 0.63 \\
R17 & Ag 0.999 & 47.00 & 3.960 & 62.26 & 0.865 & 10480 & 4033.2 & 0.019 & 202 & 2554 & 2813 & 1.6 \\
C1 & steel-like & 39.50 & 4.072 & 28.59 & 0.723 & 7921 & 9416.4 & 0.036 & 98 & 2883 & 5136 & 1.5 \\
C2 & steel-like & 39.50 & 4.197 & 28.85 & 0.703 & 7976 & 9393.9 & 0.057 & 71 & 2107 & 5114 & 1.6 \\
C3 & steel-like & 39.50 & 4.053 & 29.06 & 0.747 & 7829 & 9442.0 & 0.065 & 68 & 2008 & 5012 & 3.3 \\
C4 & steel-like & 39.67 & 4.144 & 29.21 & 0.716 & 7969 & 9458.9 & 0.052 & 87 & 2572 & 5170 & 1.8 \\
C5 & brass-like & 39.10 & 3.661 & 26.68 & 0.728 & 8340 & 5794.3 & 0.168 & 555 & 10101 & 3518 & 1.5 \\
\bottomrule
\end{tabular}
\end{table*}

\subsection{Higher modes}
A tapped disc is not a string, and the recorded overtones say so quantitatively.
Pooling 1970 higher-frequency peaks from 1207 accepted taps analysed at the
production frequency floor, \SI{83.7}{\percent} fall within \SI{2}{\percent} of a free-plate modal ratio while only
\SI{10.4}{\percent} fall within \SI{2}{\percent} of an integer multiple of $f_{1}$; the
mean distance to the nearest free-plate mode is 0.044 against 0.268 to the nearest
integer. Shifting every ratio by a random factor of up to $\pm$\SI{10}{\percent}
drops the free-plate fraction to \SI{21}{\percent}, so the agreement is not an
artefact of how densely the free-plate ratios are spaced. This is worth reporting rather than assuming, because the loose term
``harmonics'' invites the expectation of integer ratios, and a method that searched for
them would find nothing.

\begin{figure}[!htbp]
\centering
\includegraphics[width=\columnwidth]{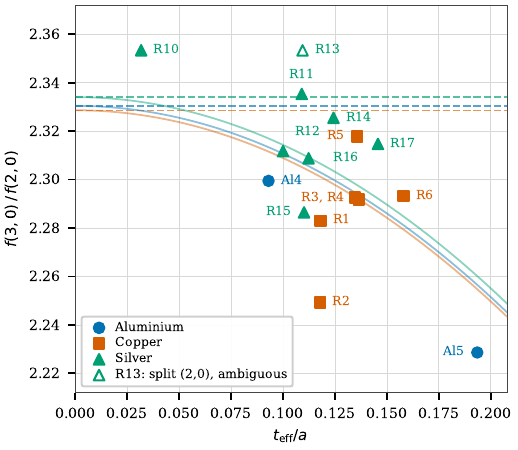}
\caption{The $(3,0)/(2,0)$ frequency ratio against aspect ratio, for all sixteen
specimens in which both modes resolve and the geometry is known. Dashed lines are the
Kirchhoff values for each material's Poisson's ratio; solid curves are a single
one-parameter fit in $(t_\mathrm{eff}/a)^{2}$ constrained to meet the Kirchhoff limit
as the plate becomes thin. Thirteen of sixteen lie below their thin-plate value, as
finite-thickness theory predicts; the $(t/a)^{2}$ trend rests mainly on the thickest
disc, Al5. The hollow marker, R13, carries a \SI{5.3}{\percent} split of its $(2,0)$ mode,
which makes its ratio ambiguous.}
\label{fig:modes}
\end{figure}

Table~\ref{tab:modes} gives the census. The dominant overtone sits at
$2.23$--$2.35\,f_{1}$, identifying it as the $(3,0)$ mode, and appears in every
specimen in which any overtone is resolved at all. A second group of peaks falls
between $3.7$ and $4.1\,f_{1}$. That group is reported as one
entry rather than two because the $(1,1)$ and $(4,0)$ modes are separated by only
\SI{2.6}{\percent} at $\nu=0.37$, against specimen-to-specimen scatter of about
\SI{1}{\percent}: the two cannot be told apart at this precision, and splitting the
count between them would claim a resolution the data does not support. The
axisymmetric $(0,1)$ mode is almost absent: a peak at \numrange{1.72}{1.77}\,$f_{1}$
appears in only 9 of 1207 accepted taps, never more than twice on any specimen.

A rim strike does not account for that rarity: $(0,1)$ has a nodal circle at
$r\approx0.68a$ and non-zero rim displacement, so a rim impulse couples to it. The
mount is the likely cause. Every other mode detected here---$(2,0)$, $(3,0)$, $(1,1)$ and $(4,0)$---has a
nodal line or circle passing through the disc centre, whereas $(0,1)$ has a centre
antinode. The centre pinch that supports the specimen without damping the modes of
interest therefore damps $(0,1)$ preferentially, and its rarity is most plausibly a
property of the fixture rather than of the excitation; this is argued rather than
demonstrated, since no specimen was re-tapped on a rim support.

\begin{table}[!t]
\caption{Higher-mode census. Predicted ratios are computed across the Poisson's ratios
of the materials studied, $\nu=0.29$ to $0.37$, and pooled over the 1207 accepted
taps of the production run.}
\label{tab:modes}
\centering
\small
\begin{tabular}{@{}lcl@{}}
\toprule
Mode & ratio to $(2,0)$ & observed\\
\midrule
$(0,1)$          & 1.67--1.79 & 9 taps of 1207\\
$(3,0)$          & 2.32--2.33 & wherever an overtone resolves\\
$(1,1)$, $(4,0)$ & 3.80--4.11 & a subset, not separable\\
\bottomrule
\end{tabular}
\end{table}

The mode structure also supplies an independent check on a contested measurement. The
machined sterling disc is thin and large, placing its fundamental at
\SI{887}{\hertz}---below the \SI{1000}{\hertz} analysis floor---so the production run
reports \SI{2087}{\hertz} instead. The ratio of the two is 2.353, against a theoretical
$(3,0)/(2,0)$ of 2.334 for silver: the mode the pipeline mistook for the fundamental is
identifiable as the $(3,0)$, which confirms the lower reading without appeal to
judgement. Re-running that specimen with the floor lowered recovers a split doublet at
886.4 and \SI{896.9}{\hertz} directly (means over the 75 taps on which both members
resolve), and re-assigns the \SI{2087.1}{\hertz} peak to the $(3,0)$ band, confirming
the inference from the raw spectrum. The same specimen yields
a $(4,0)$ at $4.104\,f_{1}$ against a predicted
4.113, agreement to \SI{0.2}{\percent} and the cleanest mode identification in the study,
because its low fundamental places every higher mode comfortably inside the recording
band.

The $(3,0)/(2,0)$ ratio is the best-determined quantity in the study, with
within-specimen standard error below \SI{0.05}{\percent}, and it is independent of
$E$, $\rho$, $a$ and $t$ \emph{in the Kirchhoff limit}. That last qualifier is where
the interest lies. The finite-thickness effects that \eqref{eq:shear} corrects act on
both modes and not equally, so at coin proportions the ratio should fall below its
thin-plate value by an amount growing with $t_\mathrm{eff}/a$. Figure~\ref{fig:modes}
plots the sixteen specimens for which both modes resolve and the geometry is known.
Thirteen lie below their material's Kirchhoff value, which is the predicted direction
(one-sided binomial sign test, $p=0.011$; the statistical units are the sixteen
specimens, each represented by its mean ratio, not the individual taps). The predicted \emph{scaling} is less secure: the departure
correlates with $(t_\mathrm{eff}/a)^{2}$ at $r=-0.64$ ($p=0.008$), but without the
thickest disc, Al5, at $r=-0.36$ ($p=0.19$), and within the copper and silver rounds
alone no trend is resolved. Struck coins of identical geometry differ in this ratio by
up to 0.034---R1 and R2 are the same design at the same size---so relief scatters the
ratio by about as much as thickness moves it.

Two of the sixteen, Al4 and Al5, are calibration discs; excluding them, eleven of the
remaining fourteen lie below their Kirchhoff value (one-sided $p=0.029$). The ratio
therefore supports the finite-thickness correction in sign, largely on specimens that
took no part in the calibration and through a quantity independent of $E$ and $\rho$,
but does not independently confirm its form. A series of machined discs between
$t/a=0.1$ and 0.2 would.

What the ratio is not is a material discriminant. The specimen means---$2.288\pm0.022$
for six copper rounds, $2.319\pm0.021$ for seven fine-silver rounds---do order
correctly by Poisson's ratio, but the magnitude rules that reading out: theory separates
copper from silver by $0.0054$, against an observed $0.031$, six times larger. Geometry
and relief dominate. And across the materials here the theoretical ratio spans only
\SI{0.6}{\percent}, less than the
observed specimen-to-specimen scatter of about \SI{1}{\percent}, so even without the
geometry term it could not separate them.

\subsection{A frequency-floor artifact}
The \SI{1000}{\hertz} peak-search floor, adopted to exclude enclosure air resonances
near \SIrange{390}{730}{\hertz}, misassigns the fundamental of any specimen ringing
below it. Exactly one specimen in this study is affected---the machined sterling
disc---and the effect is diagnosed above rather than hidden: its reported
\SI{2087}{\hertz} is its $(3,0)$ mode. No other specimen has a fundamental within
\SI{2}{\kilo\hertz} of the floor. Results for that specimen are quoted from the
lowered-floor re-run throughout.

The sterling disc is the visible case of a more general property of the estimator,
which is recorded here because it bounds where the method can be trusted. The
fundamental is taken as the \emph{lowest} peak clearing \SI{3}{\percent} of the
strongest, not the strongest peak. On a specimen whose mode dominates its spectrum
this is the safer choice, because it cannot mistake an overtone for the fundamental.
On a specimen whose mode is weak it inverts: any noise feature above the search floor
that clears \SI{3}{\percent} of that weak mode is reported as $f_{1}$ instead, and
the reported value then tracks the search floor rather than the specimen. Every
fundamental reported in this work carries a strike-to-strike scatter below
\SI{0.4}{\percent}, which a floor-tracking artifact cannot produce; that
reproducibility, rather than the peak-selection rule, is what certifies them.

\subsection{Application}
The decision rule, fixed before measurement, is: silver if non-ferromagnetic
\emph{and} Archimedes density exceeds \SI{9000}{\kilo\gram\per\cubic\meter}. All eight
silver specimens pass (\SIrange{10140}{10579}{\kilo\gram\per\cubic\meter}); all six
copper rounds carrying a displacement measurement are rejected
(\numrange{8849}{8987}), the two five-ounce rounds being excluded for want of one;
none of the five unverified rounds passes as silver: four fail on the magnet (densities
\numrange{7829}{7976}) and one on density (8340).
Because every reference specimen is non-ferromagnetic, the magnet performs no
discrimination within that set; it acts as a first filter on the unverified rounds, and
density does the remaining work.

\begin{figure}[!htbp]
\centering
\includegraphics[width=\columnwidth]{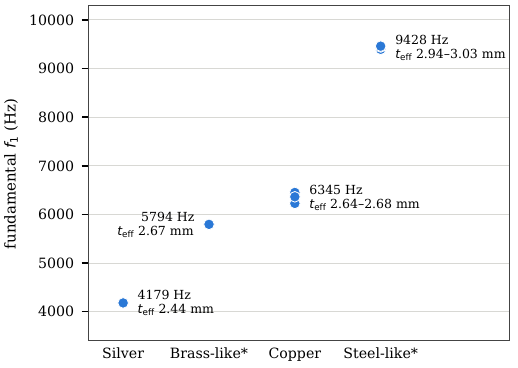}
\caption{Fundamental frequency at matched \SIrange{39.1}{39.7}{\milli\meter} diameter
for silver, copper and the rounds of unverified composition (asterisked; class inferred), with each
group's volume-equivalent thickness $t_\mathrm{eff}=\varphi t$. The four groups separate
on frequency for these specimens, which share a diameter but not a thickness or relief;
the separation does not establish frequency classification independent of geometry.}
\label{fig:matched}
\end{figure}

The acoustic readout provides an independent \emph{physical} contrast on every
call---elasticity rather than density---and constrains what the unverified rounds could
be made of: the four ferromagnetic examples return
\SIrange{5012}{5170}{\meter\per\second}, consistent with steel, and the
non-ferromagnetic example \SI{3518}{\meter\per\second}, consistent with brass. The two
readouts are not, however, statistically independent: both consume the same displaced
volume $V$, and a positive error in $V$ depresses the inferred $\rho$ and $c_{p}$
together. At matched \SIrange{39.1}{39.7}{\milli\meter} diameter the four groups also
separate on frequency alone: silver \SI{4179}{\hertz}, brass-like \SI{5794}{\hertz},
copper \SIrange{6225}{6451}{\hertz}, steel-like \SIrange{9394}{9459}{\hertz}
(Fig.~\ref{fig:matched}). Frequency alone is free of the shared-volume objection, since
it uses no displacement measurement at all. It is not, however, controlled for
geometry: these specimens share a diameter but not a thickness or a relief, and their
volume-equivalent thicknesses span \SIrange{2.44}{3.03}{\milli\meter}, a range that by
itself would move $f_{1}$ by about \SI{24}{\percent}. The separation therefore holds
for this selected group of specimens; it does not establish that frequency classifies
material independently of geometry, which is why the readout of this work is $c_{p}$
rather than $f_{1}$.

The rule fails on one specimen, and it is in the reference set. The machined tungsten
disc is non-ferromagnetic with $\rho=\SI{19487}{\kilo\gram\per\cubic\meter}$ and therefore
passes as silver. Tungsten is the standard high-density substitute precisely because it
defeats density screening. The acoustic readout is \emph{predicted} to reject it: pure
tungsten ($E=\SI{411}{\giga\pascal}$, $\rho=\SI{19250}{\kilo\gram\per\cubic\meter}$,
$\nu=0.28$) gives $c_{p}\approx\SI{4810}{\meter\per\second}$. At \SI{0.1}{\gram}
resolution the specimen's density read \SI{2.5}{\percent} below pure tungsten and
suggested a heavy alloy ($c_{p}\approx\SI{4560}{\meter\per\second}$); remeasured at
\SI{0.1}{\milli\gram} it reads \SI{1.2}{\percent} \emph{above}. That is compatible with
commercially pure tungsten given the displaced-volume systematic of
Section~\ref{sec:method}-B, and taken at face value lies above the
\SIrange{17000}{18500}{\kilo\gram\per\cubic\meter} of common tungsten heavy alloys. It
does not by itself establish the composition: no composition measurement was made, and
the same disc's displaced volume falls \SI{9.4}{\percent} short of its nominal cylinder
for reasons that cannot be uniquely attributed (Fig.~\ref{fig:volume}). Either predicted
value lies above the highest copper reading recorded here,
\SI{3972}{\meter\per\second}, so the predicted rejection does not depend on which
applies. Two attempts to test that prediction produced no acoustic reading. The disc of
Table~\ref{tab:specimens} produced no reproducibly detectable resonance on the fixture
used for the rest of this work. A second, thicker specimen was then obtained
and recorded---\SI{24.323}{\milli\meter} diameter, \SI{2.2725}{\milli\meter} thick,
\SI{15.767}{\gram}, non-ferromagnetic, with a predicted fundamental of
\SIrange{16}{24}{\kilo\hertz} depending on composition---and no resonance was detected
from it either.
Across 68 taps in five takes no peak in the predicted \SIrange{16}{24}{\kilo\hertz}
band reached \SI{15}{\deci\bel} of prominence at any analysis window between 5 and
\SI{350}{\milli\second}; the in-band maximum was \SI{14.6}{\deci\bel}. Outside that
band at most three taps per window carried an isolated peak above \SI{15}{\deci\bel},
and no frequency recurred in three of the five files within \SI{0.2}{\percent}. A candidate detection obtained by
pooling all peaks was rejected by a label-permutation null test, which reproduced it
exactly ($p=0.995$ over 200 surrogates).

That null is quantified rather than merely reported. Injecting synthetic decaying
sinusoids into the recordings' own inter-tap noise, at an amplitude calibrated to the
mode-to-impact ratio measured on the aluminium control and scaled to this specimen's own
recorded impact amplitudes, establishes the sensitivity actually achieved: \emph{a mode
anywhere in \SIrange{16}{24}{\kilo\hertz}, excited as efficiently as aluminium's and
with $Q$ above approximately 700--1000, would have been detected.} None was. The
statement is conditional: it holds for a mode in that frequency band and damping range,
excited with the assumed efficiency relative to aluminium. The non-detection therefore
does not show that the specimen lacks an elastic resonance, only that none meeting
those conditions appears in the recordings. The bound carries the recording level
inside it. Measured the same way, as the maximum of each recording, these takes peaked
at 0.03--0.06 of full scale against the aluminium control's 0.15--0.18, a level
deficit of about \SI{11}{\deci\bel}. That accounts for only part of the
\SI{19}{\deci\bel} difference in their broadband impact-to-background ratios; the
remainder is not explained by recording level.

It is a floor, not a ceiling. The specimen's density is known only geometrically,
\SI{14932}{\kilo\gram\per\cubic\meter}, and the first tungsten disc shows how far a
geometric and a displaced volume can disagree: its nominal cylinder exceeded its
displaced volume by \SI{9.4}{\percent}. Even corrected by that much the second disc would
read about \SI{16500}{\kilo\gram\per\cubic\meter}, below the heavy alloys and nearer
cemented carbide or a tungsten--copper composite. Both are sintered composites, which
can damp more strongly than a fully dense metal, and a mode below $Q\approx700$ would
be invisible to this measurement while being entirely real. The honest statement is therefore that the acoustic rejection of a
high-density substitute remains predicted rather than demonstrated, and that the
sensitivity limit of the attempt is now known. A tenfold larger mode amplitude relative
to the noise would lower that limit only by about a factor of two in the injection
tests, to $Q\approx320$--470.

\subsection{What the unverified rounds are consistent with}
The readout constrains composition without identifying an alloy, and the reason is
structural: a plated or layered round is a composite, and a composite has no single $E$, $\rho$ or
$\nu$ to recover. Table~\ref{tab:suspect} compares each round's measured density and
implied wave speed against candidate materials, each evaluated with its own $\nu$ so
that no material is presupposed. The brass candidate here is 60/40 brass, the
best-matching specific alloy; Table~\ref{tab:bands} references the same specimen to
generic brass, which is why the two deviations differ.

\begin{table}[!t]
\caption{Density and wave speed of the rounds of unverified composition compared with
candidate materials. Each
candidate is evaluated with its own Poisson's ratio, so no composition is presupposed.
At the finer resolution all four ferromagnetic rounds match plain-carbon steel more
closely in density than 430 stainless (\num{7700}). A candidate is a material the
measurements are consistent with, not an identification.}
\label{tab:suspect}
\centering
\begin{tabular}{llrr}
\toprule
Round & Closest candidate & $\Delta\rho$ & $\Delta c_{p}$\\
\midrule
C1 & mild/low-carbon steel & \SI{+0.9}{\percent} & \SI{-2.6}{\percent}\\
C2 & mild/low-carbon steel & \SI{+1.6}{\percent} & \SI{-3.0}{\percent}\\
C3 & mild/low-carbon steel & \SI{-0.3}{\percent} & \SI{-5.0}{\percent}\\
C4 & mild/low-carbon steel & \SI{+1.5}{\percent} & \SI{-2.0}{\percent}\\
C5 & 60/40 brass           & \SI{-0.7}{\percent} & \SI{-4.1}{\percent}\\
\bottomrule
\end{tabular}
\end{table}

All four ferromagnetic rounds match plain-carbon steel in density to within
\SI{1.6}{\percent} and read \SIrange{2}{5}{\percent} low in wave speed against it,
comparable to their \SIrange{1.5}{3.3}{\percent} uncertainty. That deficit is neither
evidence of layered construction nor evidence against it: their fill factors,
\numrange{0.70}{0.75}, are the lowest in the study, and the residual design effect of
Section~\ref{sec:discussion}-A is of that sign and size, but a compositional difference
from the reference steel would act the same way. The defensible statements are
therefore that the four ferromagnetic rounds have density and wave speed consistent
with plain-carbon steel, and that the non-ferromagnetic fifth has properties consistent
with a brass-like copper--zinc alloy. The measurements establish no alloy grade or
elemental composition, and cannot exclude a plated or layered construction, which would
have no single $E$, $\rho$ or $\nu$ to recover.

\section{Discussion}\label{sec:discussion}

\subsection{A residual design dependence}
\begin{figure}[!htbp]
\centering
\includegraphics[width=\columnwidth]{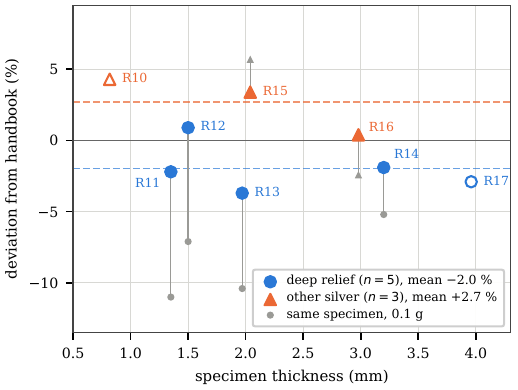}
\caption{Deviation of recovered $c_{p}$ from handbook for the eight silver specimens,
against thickness and separated by design family, at \SI{0.1}{\milli\gram} resolution.
Grey markers show the same specimens at \SI{0.1}{\gram}; dashed lines mark each
group's mean. The deep-relief family sits \SI{4.7}{\percent} below the others, with no
trend in thickness. Each specimen is
referenced to the lower bound of the accepted range for its own alloy ($E=76$\,GPa):
the machined disc against sterling, the remainder against fine silver. Hollow markers
(R10, R17) were measured only at \SI{0.1}{\gram}.}
\label{fig:design}
\end{figure}

At \SI{0.1}{\gram} resolution, sorting the silver specimens by design separated them
cleanly: five rounds of one deeply relieved commemorative design averaged
\SI{-7.4}{\percent} from handbook, three other silver specimens---a machined disc and
two shallow-relief bullion designs---averaged \SI{+2.7}{\percent}, and within the
deep-relief family the deficit shrank with thickness at $r=0.87$. Most of that was
volume resolution. At \SI{0.1}{\milli\gram} the family averages \SI{-2.0}{\percent}
and the others \SI{+2.7}{\percent} (Fig.~\ref{fig:design}); the largest single deficit
falls from \SI{11.0}{\percent} to \SI{3.7}{\percent}, and the thickness trend does not
survive ($r=-0.37$). A difference of \SI{4.7}{\percent} between the two groups remains,
at $p=0.033$ (Welch, eight specimens), but it leans on the two specimens measured only
at \SI{0.1}{\gram}: among the six measured at the finer resolution it is \SI{3.6}{\percent}
and not significant ($p=0.19$). What follows is therefore an account of an effect of
at most a few per cent, not of the \SI{11}{\percent} bias reported at the coarser
resolution.

The mechanism is modal rather than dimensional. For a plate of spatially varying
thickness $h(\mathbf{r})$ the resonance follows the Rayleigh quotient
\begin{equation}
  \omega^{2}\sim
  \frac{\displaystyle\int_{A} D(\mathbf{r})\,\mathcal{K}[w(\mathbf{r})]\,\mathrm{d}A}
       {\displaystyle\int_{A} \rho\,h(\mathbf{r})\,w^{2}(\mathbf{r})\,\mathrm{d}A},
  \qquad D(\mathbf{r})=\frac{E\,h(\mathbf{r})^{3}}{12(1-\nu^{2})},
  \label{eq:rayleigh}
\end{equation}
where $\mathcal{K}$ is the modal curvature energy density. Fixing $V$ fixes the
denominator's total mass but leaves the numerator free: what matters is where material
sits relative to modal curvature, not how much of it there is. A volume-equivalent
uniform plate therefore need not reproduce the resonance of a relieved one, and the
sign of the discrepancy is set by the relief pattern rather than by its depth alone.
A one-dimensional argument settles nothing here---averaging $h^{3}$ over the plate and
cubing the average $h$ bound the answer from opposite sides---which is why the bias is
reported as measured rather than derived.

Notably $\varphi$ itself does not order the deficit---the worst-affected specimen,
R13, has the highest fill factor in its family ($\varphi=0.91$), while R14 at
$\varphi=0.76$ is less affected---so if the effect is real it is not simply that some
coins carry more relief, but that relief interacts with the mode shape, as
\eqref{eq:rayleigh} implies.
Copper, measured over six struck rounds at $\varphi=0.78$--$0.81$, shows no such bias,
which argues against a generic failure of the correction. Those copper rounds came from
the same mint as the deep-relief silver, so the contrast also argues against a mint
effect: whatever is happening is a property of the design rather than of the
manufacturer. What the present specimen set cannot separate is design from die. All
five deep-relief specimens share one commemorative design, and no deep-relief round from
a second mint was measured, so the finding is scoped to that family rather than to
relief depth in general.

Two measurements that would settle it were outside the scope of this work. Relief depth
was never quantified---no profilometry or height mapping was performed---so the bias is
characterised by design family alone. Remeasuring R10 and R17 at the finer resolution
would decide whether the remaining group difference is significant, and a deep-relief
specimen from a second mint would then separate design from die; both are single
measurements.

The practical statement is a scope condition rather than an apology: on machined and
shallow-relief pieces the readout is unbiased to within a few per cent per specimen and
to \SI{0.9}{\percent} in the mean of six copper rounds, while on the one deeply relieved
design measured it reads \SI{2.0}{\percent} low on average and at most
\SI{3.7}{\percent}.

\subsection{The error budget points away from the acoustics}
Frequency is the best-measured quantity in the experiment, at \SI{0.15}{\percent}
between sessions. Displaced volume propagates at very nearly unity---$(q-1)/(1+q)$,
between $-0.96$ and $-1.00$ across the struck coins---and on displaced masses of
\SIrange{0.3}{6.4}{\gram} measured to \SI{0.1}{\gram} that was
\SIrange{1.6}{30}{\percent} per specimen, one to two orders of magnitude larger than
frequency. That ordering predicted that a better microphone, a quieter enclosure, or
automated excitation would improve nothing that matters, and that a finer balance would
improve everything. The prediction was tested: remeasured to \SI{0.1}{\milli\gram},
the volume term falls to \SIrange{0.5}{2.6}{\percent}, the fine-silver densities
collapse from a \SI{13}{\percent} spread (\numrange{9488}{10814}) to
\SI{0.5}{\percent} (\numrange{10530}{10579}), and the apparent design bias shrinks from
\SI{11}{\percent} to under \SI{4}{\percent}. The one physically impossible density at
the coarser resolution, \SI{10814}{\kilo\gram\per\cubic\meter} for a 0.999 silver
round, becomes \num{10575}. The binding terms are now the spread of tabulated elastic
constants ($\pm$\SIrange{0.7}{3.8}{\percent} in $c_{p}$) and, for an unknown specimen, the
choice of $\nu$ (about $\pm$\SI{3}{\percent}): neither is a measurement.

A second, irreducible term sits alongside instrument resolution. Three nominally
identical one-ounce copper rounds ring at 6225, 6450 and \SI{6361}{\hertz}, a
\SI{3.6}{\percent} spread; two of the three differ by \SI{0.5}{\percent} in displaced
mass and by \SI{1.41}{\percent} in frequency, and their recovered $c_{p}$ still spans
\SI{2.2}{\percent}. The width of any ``authentic band'' is therefore set by
manufacturing variation as much as by measurement, and acoustics cannot improve it. The advantage of an absolute readout is not a narrower
band but a \emph{predicted} rather than learned band location.

\subsection{Damping is a design signature, not a material one}
Quality factor varies nearly fivefold within a single composition class: among the
0.999 fine silver rounds $Q$ runs from 1456 to 7013 and $\tau$ from
\SIrange{0.06}{0.52}{\second}. The shallow-relief designs ring for about half a second;
the deep-relief rounds of the same metal decay in roughly a third of that. Damping is
the larger design signature: it varies fivefold with design, where $c_{p}$ varies by a
few per cent. Both plausibly follow from relief geometry acting on a mode shape, and
$Q$ is accordingly excluded from the decision rule. The same applies to splitting of the degenerate $(2,0)$ mode, which is
highly reproducible per specimen (one round gives \SI{0.653}{\percent} of $f_{1}$ with
a standard deviation of \SI{0.004}{\percent}) but measures axial asymmetry rather than
composition.

\subsection{Why no threshold on \texorpdfstring{$c_{p}$}{cp}}
Despite being the most informative single number, $c_{p}$ is deliberately not given an
acceptance threshold. The residual design effect reaches \SI{4}{\percent}; per-specimen
uncertainty spans \SIrange{0.6}{3.3}{\percent}, and \SI{6}{\percent} for the sterling
disc measured only at \SI{0.1}{\gram}; the reference values themselves carry a
half-width of \SIrange{0.7}{3.8}{\percent} from the tabulated range of $E$; and selecting
$\lambda^{2}$ requires knowing $\nu$, which for an unknown specimen is precisely the
question, contributing a further systematic of roughly $\pm$\SI{3}{\percent} across
$\nu=0.28$--$0.37$. None of this threatens the separation of silver from the non-silver
groups measured here---silver lies about \SI{15}{\percent} below the nearest of them,
the single brass-like round, and about \SI{60}{\percent} below the value handbook
constants predict for tungsten---but it does mean the readout should not be quoted to
better than a few per cent for an unknown coin.

\subsection{Comparison with prior work}
The distinction from~\cite{dao2022} and~\cite{pingcoin} is transferability rather than
accuracy. Both measure a frequency and compare it against frequencies previously
observed for that same object, so both require the coin type to be known and a library
to exist. The method reported here measures a material constant, conditioned on an
assumed Poisson's ratio, and therefore has no coverage list. The practical test is available in this data set: the reference
specimens are dominated by privately minted rounds---the class the commercial state of
the art identifies as its weakest---and the method reads them, with deep relief costing
a few per cent at most once volume is measured finely.

Relative to ultrasonic pulse-velocity testing, which is more accurate and commercially
mature, the claim is narrower: the same class of bulk elastic measurement obtained from
a tap and a microphone, without contact transducer, couplant or dedicated instrument.
The contribution is the geometry correction that makes a struck coin tractable at all.

\subsection{Limitations}
\begin{enumerate}
\item The classification has not been run blind, and this is a scope limit rather than
a pending step. The decision rule was fixed before measurement, which is necessary but
not sufficient: without specimens relabelled by a third party and classified without
knowledge of identity, no sensitivity, specificity or false-positive rate can be quoted.
What this work demonstrates is therefore a separation of known specimens, not a
validated classifier, and every claim in it is worded accordingly.
\item Tungsten defeats the density decision rule, and the acoustic rejection is
predicted, not demonstrated: neither tungsten specimen produced a reproducibly
detectable resonance under the conditions used. The injection test bounds what a mode of
given frequency, damping and excitation efficiency would have produced; it does not show
that the specimens lack a resonance. The first specimen's density is compatible with,
but does not establish, commercially pure tungsten.
\item A residual design dependence of a few per cent remains on one deeply relieved
silver design. It is significant only when two specimens measured at \SI{0.1}{\gram}
are included; relief depth was not measured; and all five affected specimens share one
design, leaving design and die confounded. The mechanism proposed in
Section~\ref{sec:discussion}-A is consistent with the observation but is not established
by it.
\item Three reference specimens (R6, R10, R17) have displaced mass measured only at
\SI{0.1}{\gram}. The two five-ounce copper rounds (R7, R8) have no displacement
measurement at all; they are outside the six-round copper validation set (R1--R6) and
every result that needs a volume. The \SI{0.5}{\percent} systematic floor applied at the
finer resolution rests on one machined disc and on the fine-silver density excess over
handbook, not on an independent calibration, and that excess may be a common bias
rather than scatter (Section~\ref{sec:method}-B).
\item The per-specimen uncertainties $u$ are measurement-derived. They exclude the
Poisson's-ratio term for an unknown material, relief model error, displacement bias
beyond the floor, and $\beta$-transfer error, and $\beta$ itself is an empirical
coefficient from a five-disc fit.
\item The compositions of C1--C5 have not been independently established. Confirming
them would require an independent bulk analysis---a chemical assay, or spectroscopy on a
cross-section, since a surface measurement cannot see beneath a plating---and that is
the natural next step for these specimens. Until then the steel-like and brass-like
labels are inferences from the measured properties only.
\item The unverified rounds' diameters were read to \SI{0.5}{\milli\meter}, which at a
sensitivity of four dominates their $c_{p}$ uncertainty (\SIrange{1.5}{3.3}{\percent});
their \SIrange{2}{5}{\percent} deficit against plain-carbon steel is of the same order.
\item The acoustic readout and the density criterion share the displaced volume $V$,
so their errors are correlated; only the matched-diameter frequency comparison of
Section~\ref{sec:results}-G is free of that coupling, and that comparison is not
controlled for thickness or relief.
\item The detection statistic is blind to heavily damped specimens. At $Q$ of 100--150
and below, the injection tests of Section~\ref{sec:method}-E produced no detection at any
amplitude up to half of full scale; the boundary depends on frequency and on the
analysis window, and was established empirically rather than derived. No specimen in this study
approaches that limit, but it bounds the method: a substitute made of a sufficiently
lossy material would return no acoustic reading at all rather than a wrong one. Density
would still reject it; the acoustic axis would simply abstain.
\item Excitation was manual throughout.
\item The analysis frequency floor misassigns the fundamental of any specimen ringing
below \SI{1}{\kilo\hertz}. One specimen in this study is affected and is diagnosed in
Section~\ref{sec:results}-F; results for it are taken from a lowered-floor re-run.
\item $\tau$ and $Q$ are extrapolations on the two most lightly damped specimens.
The decay fit window is capped at \SI{800}{\milli\second}, yet aluminium disc~4
returns a median $\tau$ of \SI{895}{\milli\second} (\SI{2.17}{\second} on its
longest-ringing tap) and the five-ounce copper round a median of
\SI{946}{\milli\second}. Both are fitted over less than one time constant---under
\SI{8}{\deci\bel} of decay---and the latter carries the largest $Q$ in the study,
12\,974. Neither should be read as a measurement.
\item Sample coverage is uneven; statements about the reference set are scoped to the
thirteen specimens with five analysed takes.
\end{enumerate}

\section{Conclusion}
A fill-factor correction, obtained from a displacement measurement, supplies the
effective thickness that relief renders unmeasurable and thereby converts an empirical
acoustic comparison into an absolute measurement of plate wave speed. Calibrated on
aluminium and validated on six struck copper rounds that took no part in that
calibration---recovering a modulus of $119.1\pm6.7$\,\si{\giga\pascal} at the centre of
an accepted \SIrange{110}{128}{\giga\pascal}---the method, using commodity audio
hardware, recovers wave speeds that form distinct observed groups across the measured
specimens, with the copper rounds providing the strongest independent validation. The
point estimates of those groups do not overlap, but the measurement uncertainties do not
resolve all six as distinct composition classes. With displaced mass measured to
\SI{0.1}{\milli\gram}, five of the six groups recover a modulus inside or at the edge
of their accepted range; only the single brass-like round falls clearly below. The
labels attached to two of those groups, the rounds of unverified composition, are inferred from density,
magnetism and wave speed rather than independently known. \emph{Absolute} here means
computed from measured resonance and geometry through a physical model rather than
retrieved from a coin-specific frequency library; it does not mean an assumption-free
determination of intrinsic material properties, since the readout is conditioned on an
assumed Poisson's ratio. The
uncertainty budget predicted that a finer balance, not better acoustics, would improve
the measurement, and remeasurement confirmed it: per-specimen uncertainty fell from
\SIrange{1.6}{30}{\percent} to \SIrange{0.5}{2.6}{\percent}, and an apparent
\SI{11}{\percent} design bias shrank to a residual of a few per cent on one deeply
relieved design. The extension that would most improve the measurement now is an
excitable tungsten specimen, which would convert the one case that defeats the density
criterion from a prediction into a demonstration. Two further extensions would broaden what the work can claim rather than
improve what it measures: a blind classification trial, and a deep-relief specimen from
a second mint to separate design from die.

\section*{Author Contributions}
S.~Dong developed the volume-based fill-factor correction for struck coins, wrote the
analysis software, performed the experimental measurements and data analysis, and
prepared the manuscript. T.~Liu
conceived and developed the acoustic measurement approach, provided the experimental
equipment and research infrastructure, and supervised the research. Both authors
contributed to the development and interpretation of the methodology, reviewed the
manuscript, and approved the submitted version.


\appendices
\section{Free-Plate Coefficients by Direct Solution}\label{app:lambda}

Leissa tabulates the completely free circular plate at two Poisson's ratios only,
nominally $\nu=0.33$ and $\nu=0.25$, and supplies no interpolation formula. The specimens here
span $\nu=0.28$ to $0.37$, so the coefficients were obtained by solving the eigenvalue
problem directly.

For a free plate the deflection of the mode with $n$ nodal diameters is
$W(r)=A\,J_{n}(kr)+C\,I_{n}(kr)$, with $k^{4}=\rho h\omega^{2}/D$. The free edge imposes
zero radial bending moment and zero Kelvin--Kirchhoff shear at $r=a$:
\begin{align}
M_{r} &= -D\left[W''+\nu\left(\tfrac{1}{r}W'-\tfrac{n^{2}}{r^{2}}W\right)\right]=0,\\
V_{r} &= -D\left[(\nabla^{2}W)'-(1-\nu)\tfrac{n^{2}}{r}\left(\tfrac{W'}{r}-\tfrac{W}{r^{2}}\right)\right]=0 .
\end{align}
Writing $x=ka$ and eliminating $A$ and $C$ gives a $2\times2$ determinant whose roots
are $\lambda^{2}=x^{2}$. With
\begin{align}
\mathcal{M}_{J} &= x^{2}J_{n}+(1-\nu)\left(xJ_{n}'-n^{2}J_{n}\right),\\
\mathcal{M}_{I} &= -x^{2}I_{n}+(1-\nu)\left(xI_{n}'-n^{2}I_{n}\right),\\
\mathcal{V}_{J} &= x^{3}J_{n}'+(1-\nu)n^{2}\left(xJ_{n}'-J_{n}\right),\\
\mathcal{V}_{I} &= -x^{3}I_{n}'+(1-\nu)n^{2}\left(xI_{n}'-I_{n}\right),
\end{align}
the frequency equation is $\mathcal{M}_{J}\mathcal{V}_{I}-\mathcal{M}_{I}\mathcal{V}_{J}=0$,
solved here by bracketing sign changes and refining by Brent's method.

The relative sign between $\mathcal{V}_{J}$ and $\mathcal{V}_{I}$ is the step at which
this derivation most easily goes wrong: reversing it leaves the $n=0$ and $n=1$ roots
correct while displacing every $n\ge2$ root, which is difficult to detect without a
reference table. Table~\ref{tab:validate} is therefore reported as a check rather than
assumed.

\begin{table}[!t]
\caption{Solver validated against both of Leissa's free-plate tables}
\label{tab:validate}
\centering
\begin{tabular}{lcccc}
\toprule
& \multicolumn{2}{c}{$\nu=0.25$} & \multicolumn{2}{c}{$\nu=1/3$}\\
\cmidrule(lr){2-3}\cmidrule(lr){4-5}
Mode & computed & Leissa & computed & Leissa\\
\midrule
$(2,0)$ &  5.511 &  5.513 &  5.251 &  5.253\\
$(0,1)$ &  8.890 &  8.890 &  9.076 &  9.084\\
$(3,0)$ & 12.744 & 12.750 & 12.222 & 12.230\\
$(1,1)$ & 20.409 & 20.410 & 20.517 & 20.520\\
$(2,1)$ & 35.290 & 35.280 & 35.241 & 35.250\\
$(0,2)$ & 38.335 & 38.340 & 38.514 & 38.550\\
\bottomrule
\end{tabular}
\end{table}

Agreement is within \SI{0.04}{\percent} against Table~2.6 and \SI{0.09}{\percent}
against Table~2.5 on the six modes Leissa tabulates at both Poisson's ratios. Leissa
heads Table~2.5 $\nu=0.33$, but the solver reproduces it only to \SI{0.17}{\percent}
at $\nu=0.33$ and to \SI{0.09}{\percent} at $\nu=1/3$; the tabulated values evidently
correspond to $\nu=1/3$, rounded in the heading, and the comparison is made there. The
$(4,0)$ mode is omitted from the comparison because Leissa quotes it to three
significant figures only. Two independent implementations were
written and agree. The coefficients of Table~\ref{tab:lambda} follow.

\section{Analysis Provenance}\label{app:provenance}

Two per-tap output files cover the same aluminium session. Both segment the audio
identically---tap counts agree file by file---yet they report different fundamentals for
91 of 214 shared taps, because $f_{1}$ is sometimes assigned to a different member of a
split doublet. The earlier file reports a single peak in the neighbourhood of the
fundamental where the later reports two to four, which is the signature of a coarser
minimum peak separation: the analysis program's minimum reported peak separation was
lowered from \SI{30}{\hertz} to \SI{8}{\hertz} between the two runs, and \SI{30}{\hertz}
merges every doublet narrower than that. The recorded command line alone does not
determine the output, since the change was to a default rather than to a flag, and the
script version is archived with each result accordingly.

The calibration is insensitive to the choice. Fitting from either file, and from the
earlier file under either doublet treatment, changes $C$ by less than \SI{0.02}{\percent}
and the recovered modulus by less than \SI{0.03}{\giga\pascal}
(Table~\ref{tab:provenance}). The five aluminium fundamentals of
Table~\ref{tab:specimens} are all drawn from the regenerated file (recordings 121--148,
dominant cluster), and re-running the final analysis program on the original recordings
reproduces that file exactly.

\begin{table}[!t]
\caption{Calibration is insensitive to the doublet treatment. The last column is a
fit for $C$ alone at fixed $\beta$, with its own standard error.}
\label{tab:provenance}
\centering
\begin{tabular}{lccc}
\toprule
Treatment & uncorr. $C$ & joint $\beta$ & $C$ at $\beta=0.895$\\
\midrule
Regenerated file        & 1182.8 & $0.895\pm0.047$ & $1292.2\pm1.9$\\
Earlier file, averaged  & 1183.0 & $0.898\pm0.048$ & $1292.4\pm1.9$\\
Earlier file, lower cl. & 1182.9 & $0.895\pm0.047$ & $1292.3\pm1.9$\\
\bottomrule
\end{tabular}
\end{table}

A related estimator choice deserves recording. For the calibration discs, specimen-level
$f_{1}$ is the mean of the \emph{dominant} cluster, not the lowest. On one disc the lowest cluster is a single stray
tap \SI{0.2}{\percent} below a population of thirty-two; taking it would bias that
specimen's fundamental by the full \SI{0.2}{\percent}, comparable to the entire
between-session reproducibility budget.

Specimen identity is taken from the laboratory take log, which records the specimen
and take number for every recording; the per-tap output carries the recording number,
so the join is exact. Five takes the log marks as scrap or dud (four on aluminium discs,
one immediately before C1) passed the automatic quality gates and remain in the reported
means; excluding them moves no fundamental by more than \SI{0.02}{\percent}. One log entry
(recording 149, logged as aluminium disc 1) disagrees with its own spectrum, which
places it unambiguously on disc 5 at \SI{7100}{\hertz}; the spectrum is followed.
Frequency alone would not have sufficed: two reference specimens, R7 and R13, differ by
only \SI{0.04}{\percent} in $f_{1}$, and the four ferromagnetic unverified rounds by
\SIrange{0.2}{0.7}{\percent}.

\bibliographystyle{IEEEtran}
\bibliography{references}

\end{document}